\documentclass[12pt]{article}
																			 																																																												\pdfoutput=1	
\usepackage{amsmath,amssymb}
\usepackage{graphicx}
\usepackage[pdftex]{hyperref}
\usepackage{epstopdf}
\usepackage{xcolor}
\DeclareGraphicsExtensions{.eps,.bmp,.wmf,.jpeg,.pdf}
\numberwithin{equation}{section}
\def\be{\begin{equation}}
\def\ee{\end{equation}}

\def\bea{\begin{eqnarray}}
\def\eea{\end{eqnarray}}
\title{$\Lambda$CDM without cosmological constant}
\author{L. N. Granda\thanks{luis.granda@correounivalle.edu.co}\\ {\small\it Departamento de Fisica, Universidad del Valle}\\{\small\it A.A. 25360, Cali, Colombia}}
\date{}
\begin{document}
\bibliographystyle{alpha}
\maketitle

\begin{abstract}
\noindent 
A type of exponential correction to General Relativity gives viable modified gravity model of dark energy. The model behaves as $R-2\Lambda$ at large curvature where an effective cosmological constant appears, but it becomes zero in flat space time.  The cosmic evolution of the main density parameters is consistent with current observations. The thin shell conditions for the Solar system were analyzed. Apart from satisfying cosmological and local gravity restrictions, the model may also show measurable differences with $\Lambda$CDM at recent times. The current value of the deviation parameter $m$ for scales relevant to the matter power spectrum can be larger than $10^{-6}$. The growth index of matter density perturbations is clearly different from that of the $\Lambda$CDM. The theoretical predictions of the model for the weighted growth rate were analyzed in the light of the $f\sigma_8$-tension.
\end{abstract}

\section{Introduction}
\noindent 
Modified gravity models have gained interest with the recent discovery of the gravitational waves that also imposed stringent restriction on the velocity of its propagation. Due to this restriction, some dark energy scalar-tensor models and models belonging to the class of Horndeski or Galilean theories \cite{horndeski, nicolis, deffayet} have been severely restricted, to the point of being discarded. Though the modified gravity models can avoid the restriction imposed by the velocity of gravitational waves, they have to pass severe restrictions mostly related with the local gravity tests. At cosmological scales they must be very close to, currently most successful dark energy model, $\Lambda$CDM (for review see \cite{copeland, sahnii, padmanabhan, sergeiod}) but locally they must reproduce with great accuracy the results of General Relativity (GR).
The function $f(R)$ that generalizes the Einstein-Hilbert Lagrangian must contain corrections that are non-linear functions of the curvature  (see \cite{sodintsov1, sotiriou, stsujikawa0, tsujikawa0, sodintsov1a, odinnojiri, odinoiko4} for reviews). 

These corrections should play an important role in the late universe and provide the necessary conditions for the transition from the decelerated to the accelerated phase of expansion, consistent with current observations, and at the same time these corrections should not be significant for local gravitational systems.
Among the variety of proposed models that cause accelerated expansion are \cite{sodintsov1, capozziello, capozziello1, sodintsov, nojiri5, carroll, nojiriodin, nojiri, elizalde, troisi, allemandi, koivisto, msami, barrow1a, faraoni, brevik, koivisto1, sotiriou1, nojiri1, dobado, sodintsov2, anthoni, nojiri2, faraoni1, song, bean, olmo, amendola1, barrow1, fay, faraoni2, hu, tsujikawa1, elizalde1, sergeioiko, sergeisaez, sergeisaez1}. Models containing positive and negative powers of curvature are among the first and most studied corrections 
to the Einstein-Hilbert Lagrangian, where it was found that corrections with positive powers of curvature are relevant at early times, like in the case of $R^2$ Starobinsky model \cite{starobinsky}, while models with negative powers of curvature, which although lead to late time accelerating universe, contain instabilities that prevent them from having a matter dominated era \cite{dolgov, lucamendola, nojiri5, sodintsov2} and are also inconsistent with Solar system tests. Attempts to unify early time inflation with late time acceleration have been considered in \cite{nojiri5, nojiri6, nojiri7, nojiri8}.
The Gauss-Bonnet 4-dimensional invariant has also been considered in the context of modified gravity in  \cite{nojiriodintsov0, nojiriodintsov1, nojiriodintsov2, cognolaelizalde}, where it was shown that some functions of the Gauss-Bonnet invariant can lead to viable cosmological solutions with accelerated expansion.  
Exact cosmological solutions have been studied in \cite{bamba1, barrow, clifton, capozz, capozz1, capozz2, barrow2} and $f(R)$ models that can satisfy both cosmological and local gravity constraints have been proposed in \cite{hu, astarobinsky, appleby1, sergeid1, sergeid2, eelizalde}. 
Cosmological scenarios resulting from various models of modified gravity have been investigated using the dynamical systems approach, which allows to find the critical points of the models that describe the different phases of evolution of the universe \cite{amendola1, tsujikawa1, souza, tsujikawa0, gsaridakis1, odinoiko2, odinoiko3, odinoiko1}.\\
Despite the large number of works devoted to explaining late time the accelerated expansion of the universe, the definitive answer to the dark energy problem is still lacking. Modified gravity models face many challenges related to having to satisfy simultaneously large scale cosmological restrictions and stability conditions while being practically indistinguishable from General Relativity at local gravity scales. Thus, most models that may be cosmologically viable produce distortions in the metric at the level of the Solar system leading to inconsistencies with observations. To satisfy Solar-system constraints some models give rise to the chameleon mechanism that arises when the curvature of a local system is very large compared to background curvature \cite{hu, tsujikawa, brax}. \\
In the present paper we continue the study of modified gravity with an exponential function of the curvature \cite{granda1, granda2}, where a new parameter  is introduced that leads to a reacher variety of viable cosmological scenarios. The model is able to account for all above discussed restrictions, apart from the stability conditions and very accurate description of the dark energy according to current observations. 
Al large curvature the model behaves as $f(R)=R-2\Lambda$ and $f(0)=0$ giving rise to the disappearance of the effective cosmological constant in the flat space-time. So the curvature effect that induces the accelerated expansion is unrelated to quantum vacuum energy in flat space-time. It is shown that the condition of stability ($f''(R)>0$) takes place during the whole cosmological evolution and even beyond the de Sitter phase. The local gravity constrains have been analyzed for the Solar system and it was found that the model satisfied the thin-shell conditions. The evolution of the growth of matter perturbations shows departures from the $\Lambda$CDM that could be observable in the near future experiments. The evolution of the weighted growth rate $f\sigma_8$ has also been discussed.\\
This paper is organized as follows. In section 2 we present the general features of the $f(R)$ models, including the equations for the background evolution. In section 3 we present the model, showing the conditions for stability and viability and some numerical study including Solar system tests. Constraints from matter density perturbations are analyzed in section 4, and some discussion is given in section 5.

\section{Field equations and constraints for $f(R)$ }
The modified gravity is described by a general action of the form
\be\label{eq1}
S=\int d^4x\sqrt{-g}\left[\frac{1}{2\kappa^2}f(R)+{\cal L}_m\right]
\ee
where $\kappa^2=8\pi G$, $f(R)$ is a function of curvature that contains the linear Einstein term and non-linear corrections to it, and ${\cal L}_m$ is the Lagrangian density for the matter component which satisfies the usual conservation equation.  
Variation with respect to the metric gives the equation of motion
\be\label{eq2}
f_{,R}(R)R_{\mu\nu}-\frac{1}{2}g_{\mu\nu}f(R)+\left(g_{\mu\nu}\Box-\nabla_{\mu}\nabla_{\nu}\right)f_{,R}(R)=\kappa^2 T_{\mu\nu}^{(m)}
\ee
where $T^{(m)}_{\mu\nu}$ is the matter energy-momentum tensor and $f_{,R}\equiv \frac{df}{dR}$. The trace of eq. (\ref{eq2})  gives 
\be\label{eq3}
Rf_{,R}(R)-2f(R)+3\Box f_{,R}(R)=\kappa^2 T^{(m)}=\kappa^2\left(3p_m-\rho_m\right)
\ee
The time and spatial components of the Eq. (\ref{eq2}) are given by the following expressions
\be\label{eq2a}
3H^2f_{,R}=\frac{1}{2}\left(Rf_{,R}-f\right)-3H\dot{f}_{,R}+\kappa^2\rho_m
\ee
and
\be\label{eq2b}
-2\dot{H}f_{,R}=\ddot{f}_{,R}-H\dot{f}_{,R}+\kappa^2\left(\rho_m+p_m\right)
\ee
where dot represents derivative with respect to cosmic time.
The equations (\ref{eq2a}) and (\ref{eq2b}) can be written in the standard form 
\be\label{back8}
3H^2=\kappa^2\left(\rho_m+\rho_{DE}\right),\;\; 2\dot{H}=-\kappa^2\left(\rho_m+\rho_{DE}+p_{DE}\right),
\ee
where 
\be\label{back9}
\rho_{DE}=\frac{3}{\kappa^2}H^2-\rho_m=\frac{3}{\kappa^2}\left[\frac{1}{6}\left(Rf_{,R}-f\right)-H^2\left(f_{,R}+R' f_{,RR}-1\right)\right]
\ee
\noindent {\bf Background evolution}\\
To solve numerically the field  equations we follow the method suggested in \cite{hu,bamba1} and use the following variables
\be\label{back1}
y_{H}=\frac{H^2}{\tilde{m}^2}-a^{-3},\;\; y_R=\frac{R}{\tilde{m}^2}-3a^{-3}
\ee
and work with the $e$-fold variable $\ln a$, where (') indicates $d/d\ln a$. Note that if we assume 
\be\label{back2}
\tilde{m}^2=\frac{1}{3}\kappa^2\rho_{m0},
\ee
then from (\ref{back8}) we can write $y_H\simeq\rho_{DE}/\rho_{m0}$, which can be interpreted as a scaled DE density (ignoring the radiation component).
 Taking into account $$R=6\left(2H^2+HH'\right)$$ and (\ref{eq2a}) we find the following equation for $y_H$ (after decoupling from $y_R$)
\be\label{back3}
y''_H+J_1y'_H+J_2 y_H+J3=0
\ee
where
\be\label{back4}
J_1=4+\frac{1-f_{,R}}{6\tilde{m}^2(y_H+a^{-3})f_{,RR}}
\ee
\be\label{back5}
J_2=\frac{2-f_{,R}}{3\tilde{m}^2(y_H+a^{-3})f_{,RR}}
\ee
\be\label{back6}
J_3=-3a^{-3}-\frac{1}{6\tilde{m}^2(y_H+a^{-3})f_{,RR}}\left[(1-f_{,R})a^{-3}+\frac{1}{3}\frac{R-f}{\mu^2}\right]
\ee
and for $y_{R}$ it is found
\be\label{back7}
y_R=3\left(y'_{H}+4y_{H}\right).
\ee
From (\ref{back8}) and (\ref{back9}) we can write the EoS of dark energy in terms of $y_H$ and $y_R$ as \cite{hu,bamba1} 
\be\label{back10}
w_{DE}=-1-\frac{1}{3}\frac{y'_H}{y_H}
\ee
The background evolution can be analyzed by solving the Eq. (\ref{back3}) numerically, which allows to find $y_H$ as function of the redshift.  \\

\noindent {\bf Stability and Cosmological Constraints.}\\
\noindent Any modified gravity model $f(R)$ must obey constraints related with the stability and the avoidance of unwanted ghosts or tachyonic degrees of freedom. The condition $f_{,R}>0$ for all $R$ is necessary to avoid changing the sign of the effective Newtonian coupling preventing ghosts instabilities. The condition $f_{,RR}>0$ is required for the stability under matter perturbations at high curvature regime. In fact the scalar particle associated with $f(R)$, dubbed scalaron with mass (in matter epoch or in the regime $M^2>>R$) \cite{amendola1, tsujikawa1, tsujikawa0}
 \be\label{mass}
 M^2\simeq \frac{1}{3f_{,RR}},
 \ee 
requires $f_{,RR}>0$ in order to avoid tachyionic behavior. \\
A useful parameter to analyze the cosmological viability, derived from $f(R)$ is
\be\label{m1}
m=\frac{Rf_{,RR}}{f_{,R}},
\ee
which along with the parameter $r$ defined as
\be\label{r1}
 {\bf r}=-\frac{Rf_{,R}}{f}
 \ee
are useful to analyze the cosmological viability of $f(R)$ models. The parameter $m$ quantifies the deviation from $\Lambda$CDM, where $m=0$ corresponds to the $\Lambda$CDM model. The viable cosmological trajectories can be depicted in a $(m,{\bf r})$ diagram obtained from the dynamical system defined for the variables
\be\label{autonomous}
x=-\frac{\dot{F}}{HF},\;\; y=-\frac{f}{6H^2 F},\;\; z=\frac{R}{6H^2}=\frac{\dot{H}}{H^2}+2,\;\; w=\frac{\kappa^2\rho_r}{3H^2F},\;\;  \Omega_m=\frac{\kappa^2\rho_m}{3H^2F},
\ee
whose analysis has been carried out in detail in  \cite{amendola1}. \\
Taking into account that $f_{,R}\approx 1$ for viable models, the following relation from (\ref{mass}) and (\ref{m1}) takes place
\be\label{effmass1}
M^2\simeq \frac{R}{3m}. 
\ee
On the other hand, this mass $M$ defines a range of the force mediated by the scalaron, which determines the Compton wavelength $\lambda_C=2\pi M^{-1}$. If $\ell$ is the typical size of a local gravitational system, then the local gravity constraints on $f(R)$ are satisfied whenever $\ell>>\lambda_C$ or $M\ell>>1$, which imply that $m<<1$  for local gravitational systems \cite{tsujikawa1}. As will be shown in this paper, besides the fact that the proposed model satisfies the local gravity constraints, it still shows interesting deviations from $\Lambda$CDM in the evolution of the growth of matter perturbations.

\section{The Model}
We consider the following model for modified gravity  
\be\label{model2}
f(R)=R-\lambda \mu^2e^{-\lambda_1\left(\frac{\mu^2}{R}\right)^{\eta}}
\ee
where the dimensionless parameters satisfy $\lambda,\;\lambda_1,\;\eta>0$. 
This model satisfies the behavior
\be\label{limmodel1}
f(R>>\mu^2) \approx R-\lambda\mu^2, \;\;\; f(0)=0
\ee
where the first approximation leads to consistency with $\Lambda$CDM, and the second equality may be interpreted as the disappearance of the effective cosmological constant in the flat space-time limit. This correction to the Einstein gravity is given in the form of convergent series of negative powers of curvature. Negative powers of curvature may appear from some compactification of string/$M$-theory as was shown in \cite{sodintsov}. However, it is known that curvature corrections of the form $R^{-n}$, $n>0$ can accelerate the expansion but lead to non-standard evolution of matter era and are unstable under matter perturbations, among other problems \cite{dolgov}. As we show bellow, the model not only pass all the consistent restrictions and leads to appropriate matter and dark energy dominance eras,  but also shows deviations from $\Lambda$CDM in the evolution of matter perturbations.\\
The stability condition $f_{,R}>0$ leads to
\be\label{f1stability-2}
\frac{\mu^2}{R}<\left(\frac{1}{\eta\lambda\lambda_1}\right)^{1/(\eta+1)}
\ee
and $f_{,RR}>0$ leads to 
\be\label{f2stability-2}
\frac{\mu^2}{R}<\left(\frac{\eta+1}{\eta\lambda_1}\right)^{1/\eta}.
\ee
If $\eta\lambda_1<1$, $\eta\lambda\lambda_1<1$, then both inequalities are always satisfied as long as $\mu^2<R$. It can be shown that the simultaneous fulfillment of both inequalities takes place for a wide interval of parameter values. One can for example define a critical value for $\lambda$ such that the r.h.s of (\ref{f1stability-2}) and (\ref{f2stability-2}) are equal, which takes place for 
\be\label{lambda-c}
\lambda_c=\frac{1}{\eta+1}\left(\frac{\eta\lambda_1}{\eta+1}\right)^{1/\eta},
\ee
then if $\lambda<\lambda_c$, the stability conditions $f_{,R}>0$ and $f_{,RR}>0$ are satisfied if (\ref{f2stability-2}) takes place (i.e.  $f_{,RR}>0$) and in case $\lambda>\lambda_c$ both stability conditions are satisfied if (\ref{f1stability-2}) takes place (i.e. $f_{,R}>0$).
This correction the Einstein term can be expanded in powers in the case $\eta<1$ and $\lambda_1<<1$, since in this case the series converges rapidly. The expansion can also be performed in the case $\eta>1$, $\mu^2<R$ and $\lambda_1<1$ (in fact $\lambda_1$ can be greater than $1$ depending on the value of $\eta$)
\be\label{power-exp}
f(R)\approx R-\lambda\mu^2\left(1-\lambda_1 \left(\frac{\mu^2}{R}\right)^{\eta}\right).
\ee
This expansion is the same as that presented in the HS \cite{hu} and Starobinsky \cite{astarobinsky} models. In fact, numerical analysis shows the same results for the model (\ref{model2}) and the HS model with a suitable choice of parameter values. From (\ref{power-exp}) follows that $\lambda\mu^2$ must be very close to the observed value of the cosmological constant, i.e. $\lambda\mu^2\approx 2\Lambda$.\\
Considering formally the analytic behavior as function of curvature, the model (\ref{model2}) shows some differences with the HS model in the limit $R\to 0$: 
in the HS $f_{,R}$ diverges in the asymptotic limit ($R\to 0$) for $n<1/2$ and can eventually change the sign, while in the model (\ref{model2}) we find that $\lim_{R\to 0}f_{,R}=0$ for any $\eta>0$. The second derivative $f_{,RR}$ in the HS model also diverges for $n<2$ and in general, for the HS model, the $k$-th derivative  $|f^{k}(R)|\to \infty$ at $R\to 0$ for $n<k$. For model (\ref{model2}) all high derivatives meet the limit $\lim_{R\to 0}f_{,RR},\; f_{,RRR},... =0$ for any $\eta>0$.\\
Using $\lambda$ to fix the de Sitter curvature from ${\bf r}=-2$ at $R=R_{ds}$ we find ($R_{ds}=\mu^2 y_{ds}$)
\be\label{lambda-ds1}
\lambda=\frac{y_{ds}e^{\lambda_1\left(\frac{1}{y_{ds}}\right)^{\eta}}}{2-\eta\lambda_1\left(\frac{1}{y_{ds}}\right)^{\eta}}.
\ee
Under the condition $\lambda_1/y_{ds}^{\eta}<<1$ this expression can be kept very close to the limit 
\be\label{lambda-ds2}
\lambda\simeq\frac{1}{2}y_{ds},
\ee
Replacing $\lambda$ in (\ref{m1}) and (\ref{r1}) we find  ($R=\mu^2 y$)
\be\label{my1}
m=\frac{\eta \lambda_1y_{ds}e^{\lambda_1\left(\frac{1}{y_{ds}}\right)^{\eta}}\left(1+\eta-\eta\lambda_1 y^{-\eta}\right)}{e^{\lambda_1\left(\frac{1}{y}\right)^{\eta}}y^{\eta+1}\left(2-\eta\lambda_1 y_{ds}^{-\eta}\right)-\eta\lambda_1 y_{ds}e^{\lambda_1\left(\frac{1}{y_{ds}}\right)^{\eta}}},
\ee
\be\label{ry1}
{\bf r}=\frac{\eta \lambda_1y_{ds}e^{\lambda_1\left(\frac{1}{y_{ds}}\right)^{\eta}}- e^{\lambda_1\left(\frac{1}{y}\right)^{\eta}} y^{\eta+1}\left(2- \eta\lambda_1 y_{ds}^{-\eta}\right)}{e^{\lambda_1\left(\frac{1}{y}\right)^{\eta}}y^{\eta+1}\left(2-\eta\lambda_1 y_{ds}^{-\eta}\right)- y_{ds}e^{\lambda_1\left(\frac{1}{y_{ds}}\right)^{\eta}}y^{\eta}}.
\ee
From the condition of stability at de Sitter point, i.e. $0<m({\bf r}=-2)\le 1$, and assuming $\eta>0$, we find the following restriction on $\lambda_1$
\be\label{ds-stability}
\lambda_1\le \frac{\eta+3}{2\eta}\left[1-\sqrt{1-\frac{8}{(\eta+3)^3}}\right]y_{ds}^{\eta},\;\; {\it or} 
\ee
\be\label{ds-stability1}
\frac{\eta+1}{\eta}y_{ds}^{\eta}<\lambda_1\le  \frac{\eta+3}{2\eta}\left[1+\sqrt{1-\frac{8}{(\eta+3)^3}}\right]y_{ds}^{\eta}
\ee
This last inequality leads to negative values of $m$ in the region $y>y_{ds}$ and we discard it. Using the above discussed condition, $\lambda\mu^2\approx 2\Lambda$, and the result (\ref{lambda-ds2}) we can find an appropriate value for $y_{ds}$ as
\be\label{yds}
y_{ds}=\frac{R_{ds}}{\mu^2}\simeq \frac{4\Lambda}{\mu^2}.
\ee 
From the density parameter for the cosmological constant $\Omega_{\Lambda}$ we have $\Lambda=3H_0^2\Omega_{\Lambda}$. Taking for the mass scale $\mu^2$ the value
\be\label{mu}
\mu^2=\frac{\kappa^2\rho_{m0}}{3}=\Omega_{m0}H_0^2,
\ee
where $\rho_{m0}$ is the average current matter density, we find 
\be\label{yds1}
y_{ds}\approx 12\Omega_{\Lambda}/\Omega_{m0},
\ee
valid under the conditions for the compliance with (\ref{lambda-ds2}). Numerical analysis based on above results ((\ref{mu}), (\ref{yds1})) shows that in the case
$\eta<1$, in order to satisfy cosmological restrictions is enough to consider values of $\eta\sim 10^{-3}-1$ and $\lambda_1\sim 1$ but the corresponding values of $m$ can not satisfy the Solar system restrictions unless either $\lambda_1<<1$, taking values of the order $\lambda_1\sim 10^{-16}$ or less, or assuming $\eta\sim 10^{-18}$ or less if $\lambda_1\sim 1$. In any case, if the model satisfies simultaneously cosmological and local gravity constraints with $\eta,\lambda_1<1$, then it becomes indistinguishable from the 
$\Lambda$CDM with current time deviation parameter $m<<10^{-6}$, which is not striking. More interesting are the models with $\eta>1$, where in addition to satisfying cosmological and local gravity constraints, the model  
can show measurable deviations from $\Lambda$CDM in the evolution of the growth of matter perturbations. In Table I we present some values of the deviation parameter $m$, regarding the cosmological and local gravity constraints, for a wide range of $\eta$ and $\lambda_1$.\\

\begin{center}
\begin{tabular}{lccccc}\hline\hline
 $\eta$   &   $\lambda_1$    &$m(y_s)$   &$m(y_0)$   \\
\hline
 $10^{-4}$ &$1$ & $4.2\times 10^{-10}$ & $3.5\times 10^{-5}$\\ 
 $10^{-18}$& $1$& $ 4.2\times 10^{-24}$ & $3.5\times 10^{-19}$\\
 $10^{-2}$ &$10^{-16}$ & $3.6\times 10^{-24}$ & $3.4\times 10^{-19}$\\
$1$& $1$& $ 2.6\times 10^{-12}$ & $0.018$\\
  $2$ & $1$ & $2.3\times 10^{-18}$ & $1.3\times 10^{-3}$\\ 
 $3$ & $1$ & $1.4\times 10^{-24}$ & $6.6\times 10^{-5}$ \\ 
 $4$ & $1$ & $6.8\times 10^{-31}$ & $2.7\times 10^{-6}$ \\ 
$5$ & $1$  & $3\times 10^{-37}$ & $10^{-7}$\\ 
 $3$ & $10^2$ & $1.4\times 10^{-22}$ & $6.6\times 10^{-3}$ \\ 
$5$ & $10^4$  & $3.1\times 10^{-33}$ & $10^{-3}$\\ 
 \hline
\end{tabular}
\end{center}
\begin{center}
{\bf Table I}
\end{center}
\begin{center}
{\it Some numerical values for the parameter $m=\frac{Rf_{,RR}}{f_{,R}}$ for the Solar system ($m(y_s)$) and at current epoch ($m(y_0)$) for a wide range of $\eta$ and $\lambda_1$, where we have used $\mu^2\approx 0.3H_0^2$ and $R_s\approx  10^6 H_0^2$ for the Solar system. }
\end{center}
\subsection*{Solar-System Constraints}

\noindent The formal limit in (\ref{limmodel1}) reflects the analytical properties of the function $f$, but care must be taken when interpreting the approximation (\ref{power-exp}) which loses its validity at very small curvature where $f(R)$ goes to zero. However, even in the current low-curvature universe the expression (\ref{power-exp}) is a good approximation since $R/\mu^2$ is still large ($R/\mu^2\sim 40$), which follows assuming  (\ref{back2}) for $\mu^2$ (i.e. $\mu^2=\Omega_{m0}H_0^2$). \\ 
To analyze the effect of the model on the Newton law we can resource to the Einstein frame through the known scale transformation with 
\be\label{conformal1}
f_{,R}=e^{-\sqrt{\frac{2}{3}}\phi/M_p},
\ee
which introduces the scalar field potential   
\be\label{conformal2}
V(\phi)=\frac{M_p^2}{2}\frac{Rf_{,R}-f(R)}{f_{,R}^2}
\ee
where the scalar field $\phi$ appears coupled to the matter. In order to avoid large corrections to the Newton law induced by this scalar field, its mass $m_{\phi}$ must be large enough. Using (\ref{conformal1}) and (\ref{conformal2}) gives for this mass
\be\label{conformal3}
m_{\phi}^2=\frac{d^2V(\phi)}{d\phi^2}=\frac{1}{3}\left[\frac{1}{f_{,RR}}-\frac{4f}{f_{,R}^2}+\frac{R}{f_{,R}}\right].
\ee
Assuming the expansion (\ref{power-exp}) and keeping the leading power in $R/\mu^2$ we find
\be\label{conformal4}
m_{\phi}^2\approx \frac{\mu^2}{3\eta(\eta+1)\lambda\lambda_1}\left(\frac{R}{\mu^2}\right)^{\eta+2}
\ee
Applied to our galaxy (G) where the curvature is of the order of $R_G\sim 10^{-60} eV^2$ and therefore $R_G/\mu^2\sim 10^6$, we find the following results for $m_{\phi}$:  $\eta=5, \lambda_1=10^4$ give $m_{\phi}\sim 10^{-14} eV$, which corresponds to a Compton length $\lambda_G\sim 10^7 m$. $\eta=7, \lambda_1=10^7$ give $\lambda_G\sim 10^2$m, and for $\eta=9, \lambda_1=10^9$ it gives $\lambda_G\sim 10^{-3} m$. All these lengths are smaller than the galactic size, and therefore the correction to the Newton law is very small. All the above results improve with increasing $\eta$, where the values of $\lambda_1$ have been chosen in such a way that the model presents appreciable deviations (distinctive signals) from the $\Lambda$CDM model in the evolution of the growth of matter perturbations. 
Otherwise, for  smaller $\lambda_1$, it will be indistinguishable from $\Lambda$CDM both in the background evolution and in the effects on the growth of matter perturbations (see table I and section 4).\\
For large curvatures, typical of high-density regions, the corrections to the Newton law may become even much smaller. For experiments on the earth, where the air density $\rho\sim (10^{3} eV)^4$ gives $R\sim 10^{-44} eV^2$. Hence $R/\mu^2\sim 10^{22}$ and for the case $\eta=5, \lambda_1=10^4$ we find $m_{\phi}\sim 10^{40} eV$, making the correction to Newton's law unobservable. Similar results take place in the Jordan frame since the curvaton mass, in the approximation given in (\ref{mass}), is of the same order of magnitude as that defined in (\ref{conformal4}).\\
The behavior of the scalar field $\phi(r)$ in the Solar system (for spherically symmetric distribution of matter we will use the radial coordinate $r$) gives important information about the strength of the force mediated by this field and the post-Newtonian parameter $\gamma$, which can be used to test a given $f(R)$ model.\\ 
In the Solar system we have a spherically symmetric object of radius $r_S$ and mass $M_S$ surrounded by background matter at much lower density. We will assume that the spherically symmetric body has constant density $\rho_S$ (for $r<r_S$) and outside the body ($r>r_S$) the density is $\rho_B$, that satisfies $\rho_B<<\rho_S$. The gravitational potential on the surface of the body is given by $\Phi_S=GM_S/r_S$, where $M_S=(4/3)\pi r_S^3 \rho_S$. On the other hand, in the Einstein frame the scalar field couples to the matter Lagrangian giving rise to an effective potential of the form \cite{amanda1, amanda2, tegmark}
\be\label{solar1}
V_{eff}(\phi)=V(\phi)+e^{\beta\phi/M_p}\rho
\ee
where $\rho$ is the matter density in the Einstein frame,  $\beta=1/\sqrt{6}$ is a constant universal coupling between matter and the scalaron $\phi$ that originates in the conformal transformation, and $V(\phi)$ is given by (\ref{conformal2}). This effective potential evolves in two different density environments, presenting two different minima at the field values denoted as $\phi_S$ and $\phi_B$, i.e.
\be\label{solar2}
V'(\phi_S)+\frac{\beta}{M_p}e^{\beta\phi/M_p}\rho_s=0
\ee
\be\label{solar3}
V'(\phi_B)+\frac{\beta}{M_p}e^{\beta\phi/M_p}\rho_B=0
\ee
It was shown in \cite{amanda1,amanda2} that this potential generates the chameleon mechanism, where the mass $m_S^2=V''(\phi_S)$ in the region of high density is heavier than the mass $m_B^2=V''(\phi_B)$ in the region of lower density. It was found in \cite{amanda1} that under the chameleon mechanism a thin shell of thickness $\Delta r_S$ is formed with thin shell parameter given by 
\be\label{solar4}
\frac{\Delta r_S}{r_S}=\frac{\phi_B-\phi_S}{6\beta M_p\Phi_S},
\ee
where $\frac{\Delta r_S}{r_S}<<1$ in the thin shell regime (which suppresses the Yukawa profile in the external solution of the scalar field \cite{amanda1, amanda2, tegmark}). 
To write the expression for the thin shell parameter for the model (\ref{model2}) we need to find the fields $\phi_S$ and $\phi_B$ from the Eqs. (\ref{solar2}) and  (\ref{solar3}).  In terms of the scalar field and taking into account that $\mu^2<<R$ (in fact this condition takes place even in the current universe, where $R_0\sim (10^{-33} eV)^2$, under  the assumption (\ref{mu})  for $\mu^2$, that leads to $R_0/\mu^2\sim 40)$), the effective potential for the model (\ref{model2}) takes the form
\be\label{solar5}
V_{eff}=\frac{1}{2}\lambda \mu^2 M_p^2 e^{2\sqrt{\frac{2}{3}}\phi/M_p}\left[1-\lambda_1(\eta+1)\left(\sqrt{\frac{2}{3}}\frac{1}{\lambda\lambda_1\eta}\frac{\phi}{M_p}\right)^{\frac{\eta}{\eta+1}}\right]+e^{\frac{1}{\sqrt{6}}\phi/M_p}\rho
\ee
where, depending on the environment, $\rho=\rho_S,\rho_B$ for  $r<r_S, r>r_S$ respectively. The two minima of this potential are reached at the field values
\be\label{solar6}
\phi_S=\sqrt{\frac{3}{2}}\eta\lambda\lambda_1 M_p\left(\frac{\mu^2 M_p^2}{\rho_S}\right)^{\eta+1},\;\; r<r_S
\ee
\be\label{solar7}
\phi_B=\sqrt{\frac{3}{2}}\eta\lambda\lambda_1 M_p\left(\frac{\mu^2 M_p^2}{\rho_B}\right)^{\eta+1},\;\; r>r_S
\ee
where we used the approximation $\phi_{S,B}<<M_p$ and $\lambda<<\rho_{S,B}/(\mu^2M_p^2)$ (which is satisfied for $\lambda$ given in  (\ref{lambda-ds1}) and  (\ref{lambda-ds2}), and $y_{ds}$ given in (\ref{yds1})). Since $\rho_B<<\rho_S$, then $\phi_B>>\phi_S$ and we find the following approximation for the thin shell parameter for the model (\ref{model2})
\be\label{solar8}
\frac{\Delta r_S}{r_S}\approx \frac{1}{2}\eta\lambda\lambda_1\left(\frac{\mu^2 M_p^2}{\rho_B}\right)^{\eta+1}\frac{1}{\Phi_S}= \frac{1}{2}\eta\lambda\lambda_1\left(\frac{\mu^2 M_p^2}{\rho_B}\right)^{\eta+1}\frac{8\pi M_p^2 r_S}{M_S}.
\ee
Note from the last equality that the more massive is the object, the easier is to satisfy the thin shell condition. (similar results have been obtained in \cite{tsujikawa} for the HS and Starobinsky models).\\
The bound on the thin shell parameter can be derived from the experimental tests of the post-Newtonian parameter $\gamma$ in the Solar system, whose current tightest constraint \cite{will} is $|\gamma-1|<2.3\times 10^{-5}$. 
To find this bound we will use the condition $\phi<<M_p$, which allows as to use the approximation $r=r_{EF}\approx r_{JF}$, where $r_{EF}$ ($r_{JF}$) is the radial distance in the Einstein (Jordan) frame. Then, writing the spherically symmetric metric (the Schwarzschild metric in the weak field approximation) in the JF ($A(r),B(r)<<1$)
\be\label{solar9}
ds^2=-\Big[1-2A(r)\Big]dt^2+\Big[1+2B(r)\Big]dr^2+r^2 d\Omega^2,
\ee
it was shown in \cite{tegmark} that, under the chameleon mechanism, the post-Newtonian parameter $\gamma=B(r)/A(r)$ can take the approximate value
\be\label{solar10}
\gamma\approx \frac{1-\Delta r_S/r_S}{1+\Delta r_S/r_S},
\ee
which was obtained under the condition $\lambda_B\sim m_B^{-1}>>r_S$ ($m_B^2=V''_{eff}(\phi_B)$). In the metric (\ref{solar9}) $A(r)$ and $B(r)$ are given by the expressions (under the condition  $m_B^{-1}>>r_S$) \cite{tegmark}
\be\label{solar11}
A(r)=\frac{GM_S}{r}\left(1+\frac{\Delta r_S}{r_S}\right),\;\; B(r)=\frac{GM_S}{r}\left(1-\frac{\Delta r_S}{r_S}\right)
\ee
Using (\ref{solar10}), the experimental restriction on $\gamma$ \cite{will} leads to 
\be\label{solar12}
\frac{\Delta r_S}{r_S}<10^{-5}
\ee
Therefore, using the result (\ref{solar8}) for the model (\ref{model2}), this bound leads to 
\be\label{solar13}
\frac{1}{4}\eta y_{ds}\lambda_1\left(\frac{\mu^2 M_p^2}{\rho_B}\right)^{\eta+1}<10^{-11},
\ee
where we used the result (\ref{lambda-ds2}) for $\lambda$ and the value $\Phi_S\sim 10^{-6}$ for the Sun. Assuming (\ref{mu}) for $\mu^2$, and for the homogeneous density of baryonic and dark matter in our galaxy $\rho_B\approx 10^{-24} g/cm^3$, gives $\mu^2M_p^2/\rho_B\approx 3\times 10^{-6}$. Taking the cases ($\eta=3, \lambda_1=10^2$), ($\eta=5, \lambda_1=10^4$), ($\eta=7, \lambda_1=10^7$) give respectively $\Delta r_S/r_S\approx 1.7\times 10^{-13}, 2.5\times 10^{-22}, 3.2\times 10^{-30}$, which are much smaller than the bound (\ref{solar12}) and deviations from GR become highly suppressed. \\
Note that the spherical symmetry solution (\ref{solar9}) may include the term  $Cr^2$ added to $2A(r) $ and $2B(r)$ \cite{olmo,kamionkowski}, which takes into account that the vacuum state corresponds to the asymptotic de Sitter attractor of the model (\ref{model2}) (see Eq. (\ref{lambda-ds1})) \cite{granda1,granda2}. This term is not relevant for local tests since $C\sim H_0^2$ and therefore $H_0 r_S<<1$.
\subsection*{Background Evolution}
\noindent To analyze the cosmological evolution of the dark energy in the epoch of acceleration era we use the Eq. (\ref{back3}), considering the following initial conditions for a given redshift $z_i$ in the matter dominated era 
\be\label{ini}
y_H\Big|_{z_i} =\frac{\Omega_{\Lambda}}{\Omega_{m0}},\;\; \frac{dy_H}{dz}\Big|_{z_i} =0,
\ee
which appear naturally from the expressions (\ref{back1}) and (\ref{back7}) and using the fact that at high redshift the model is very close to the $\Lambda$CDM model. In Fig. 1 we show numerical solutions for the evolution of the scaled DE density, DE equation of state, the DE density parameter and the ratio $H^2/H_{\Lambda}^2$ showing that the background evolution of the model for $\eta>1$ is very close to that of the $\Lambda$CDM model. Nevertheless, as will be seen below, the evolution of matter perturbations show characteristics of the model that differentiate it from $\Lambda$CDM. In Fig. 2 we show numerical solutions for some cases with $\lambda_1\ne 1$
\begin{figure}
\begin{center}
\includegraphics[scale=0.57]{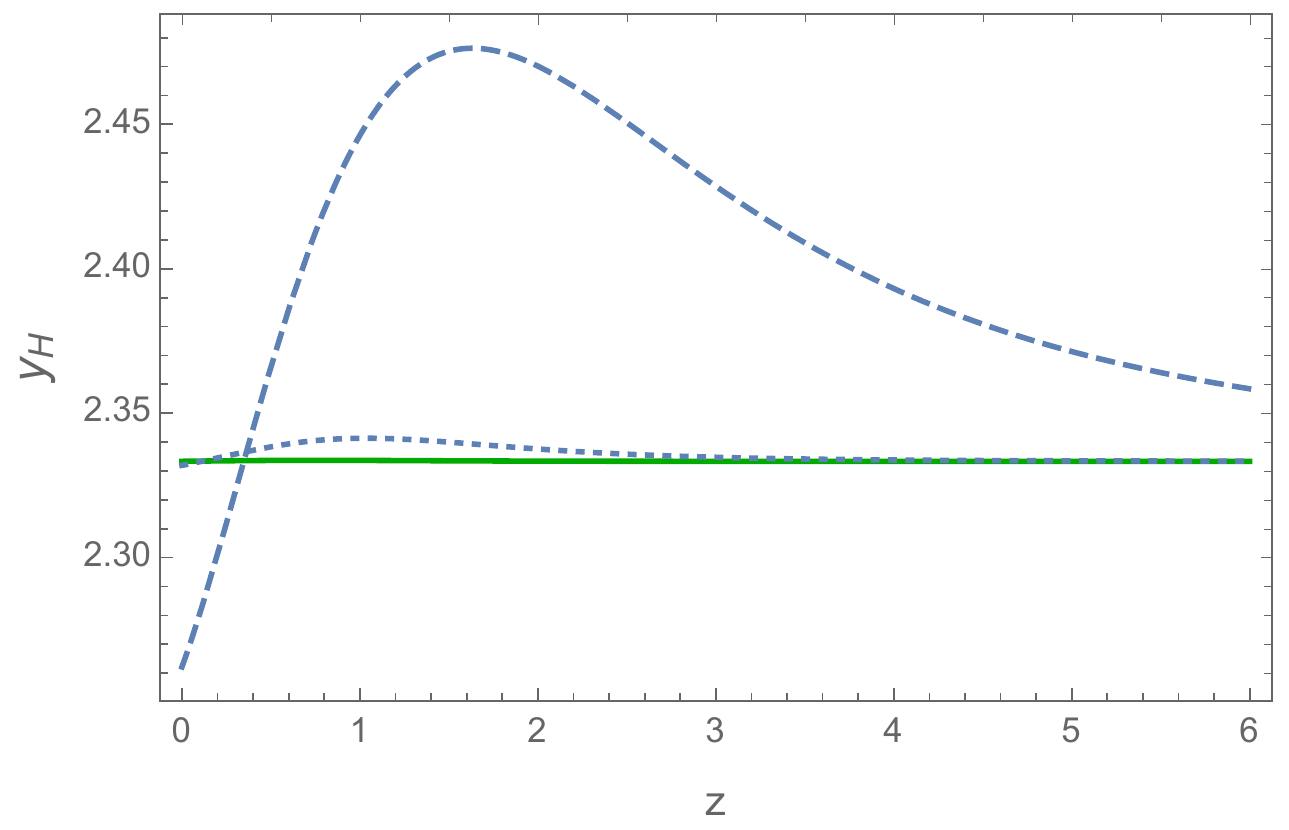}
\includegraphics[scale=0.57]{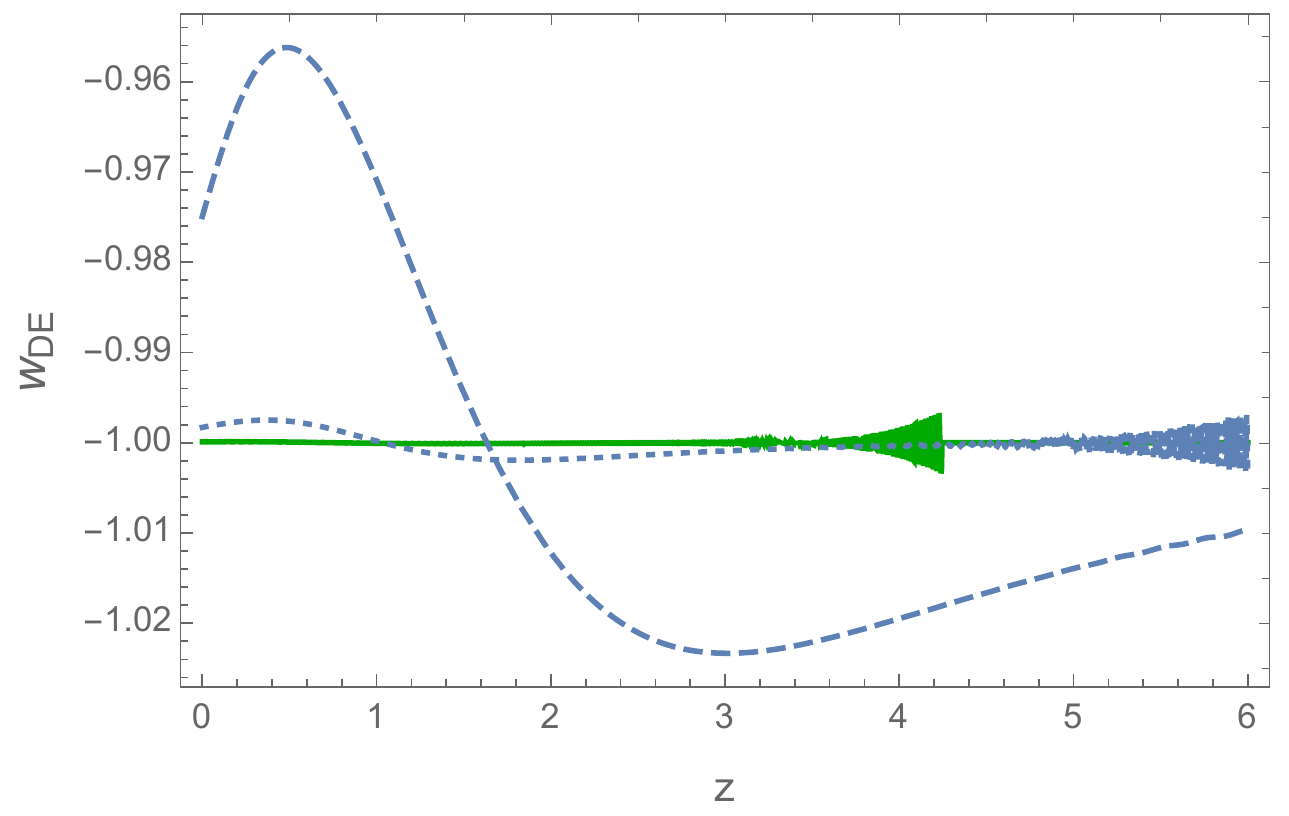}
\includegraphics[scale=0.57]{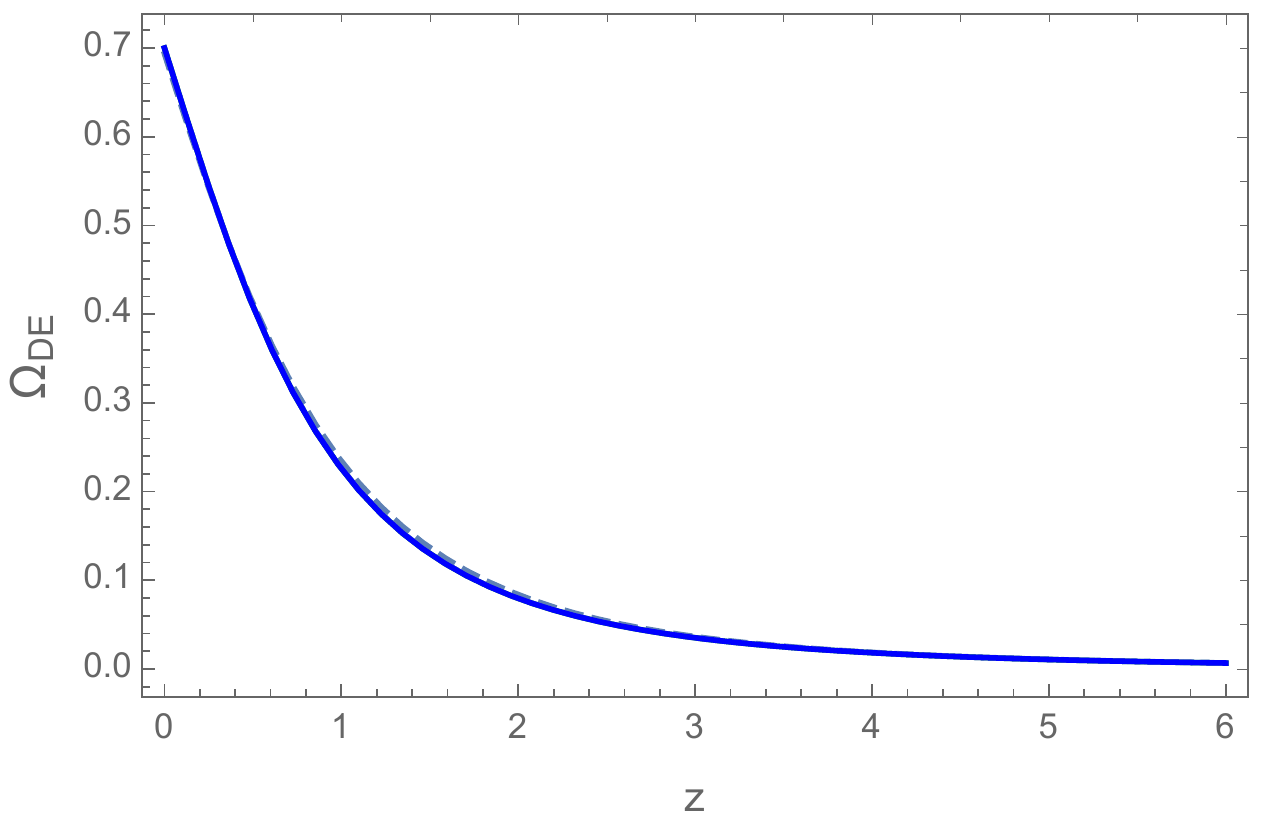}
\includegraphics[scale=0.57]{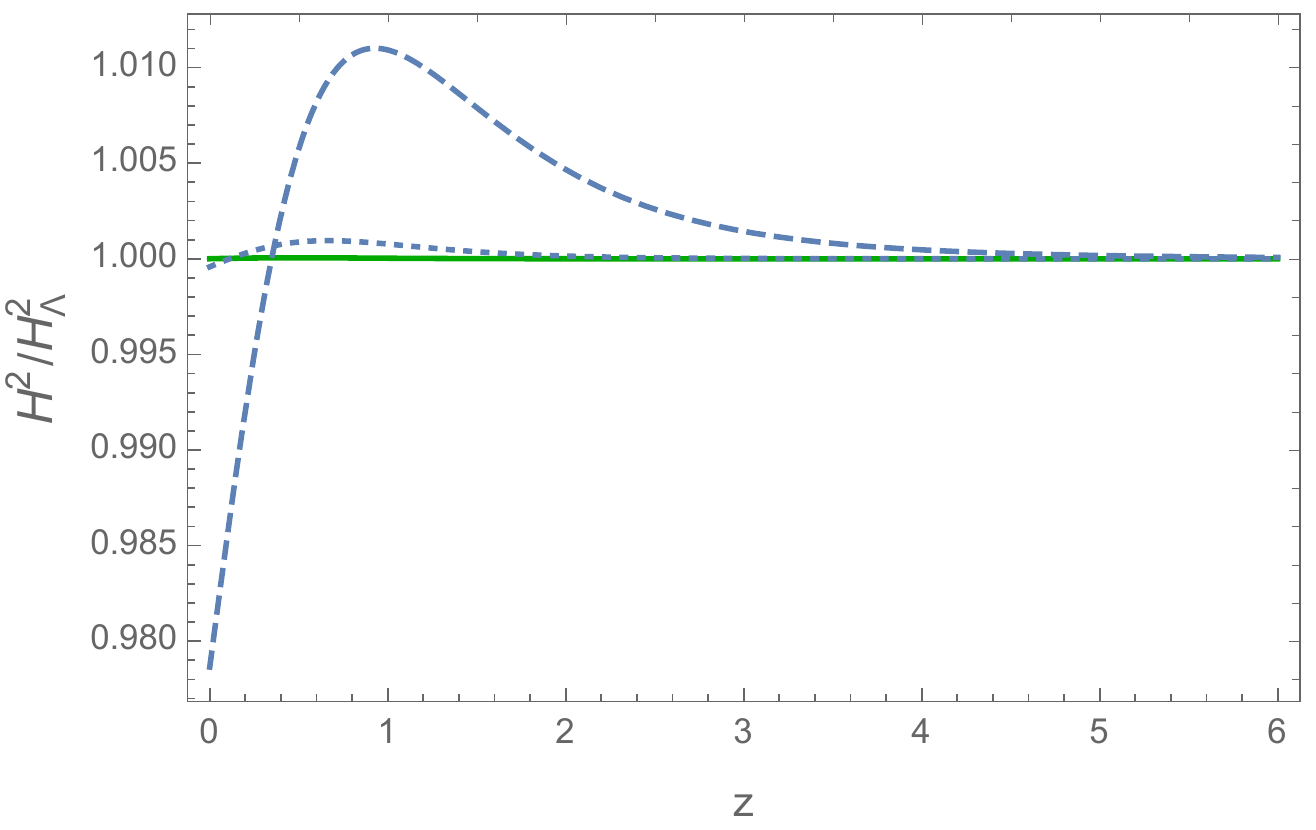}
\caption{The evolution of $w_{DE}$ and the quotient between the Hubble parameter of the model  (\ref{model2}) and that of the $\Lambda$CDM$, H^2/H_{\Lambda}^2$, assuming $\Omega_{m0}=0.3$ and  for $\lambda_1=1$. The initial redshift $z_i=19$ for $\eta=1$ (dashed), $z_i=7.7$ for $\eta=2$ (dotted) )and $z_i=98$ for $\eta=3$ (green). As $\eta$ increases, the background evolution gets closer to the $\Lambda$CDM model. In all cases the evolution of $\Omega_{DE}$ is indistinguishable from that of $\Lambda$CDM. A very narrow difference can be seen in $ H^2/H_{\Lambda}^2$.} 
\label{fig1}
\end{center}
\end{figure}

\begin{figure}
\begin{center}
\includegraphics[scale=0.57]{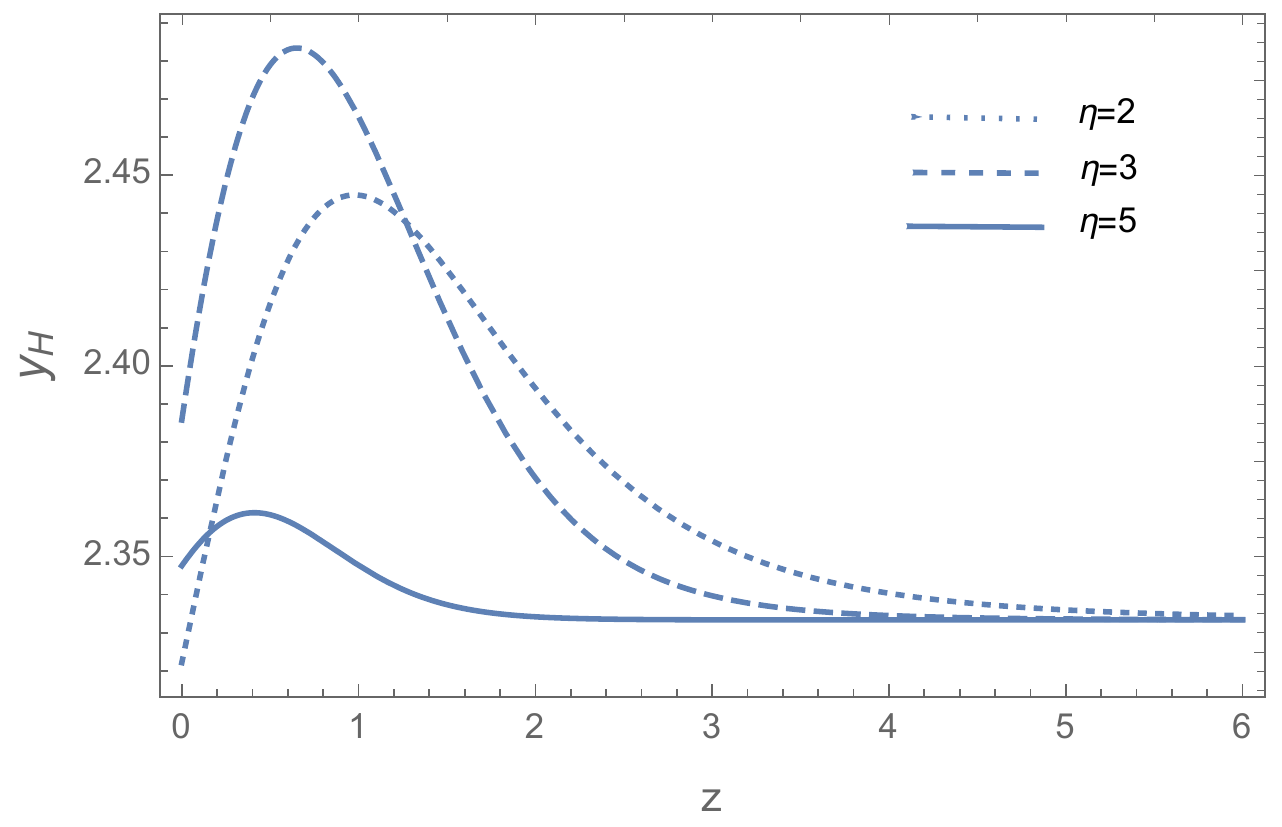}
\includegraphics[scale=0.57]{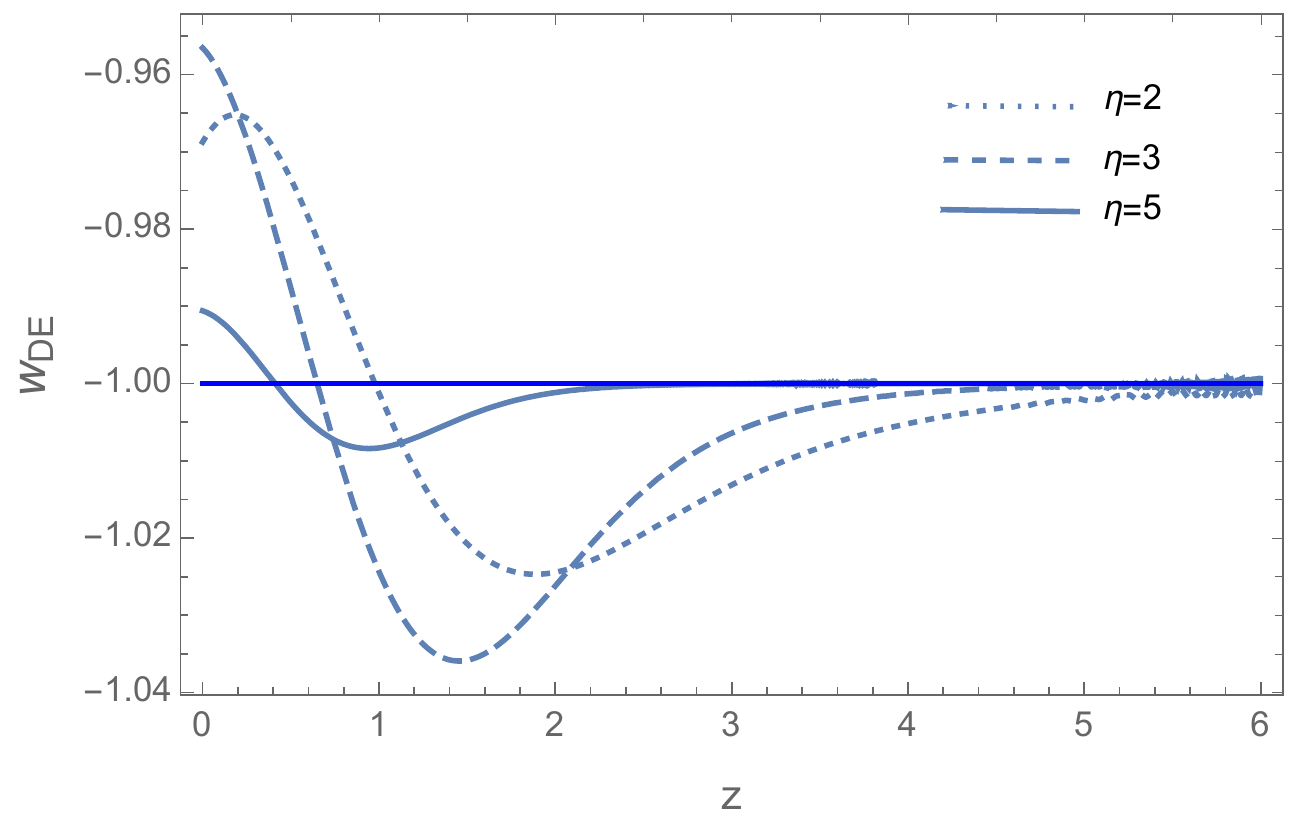}
\includegraphics[scale=0.57]{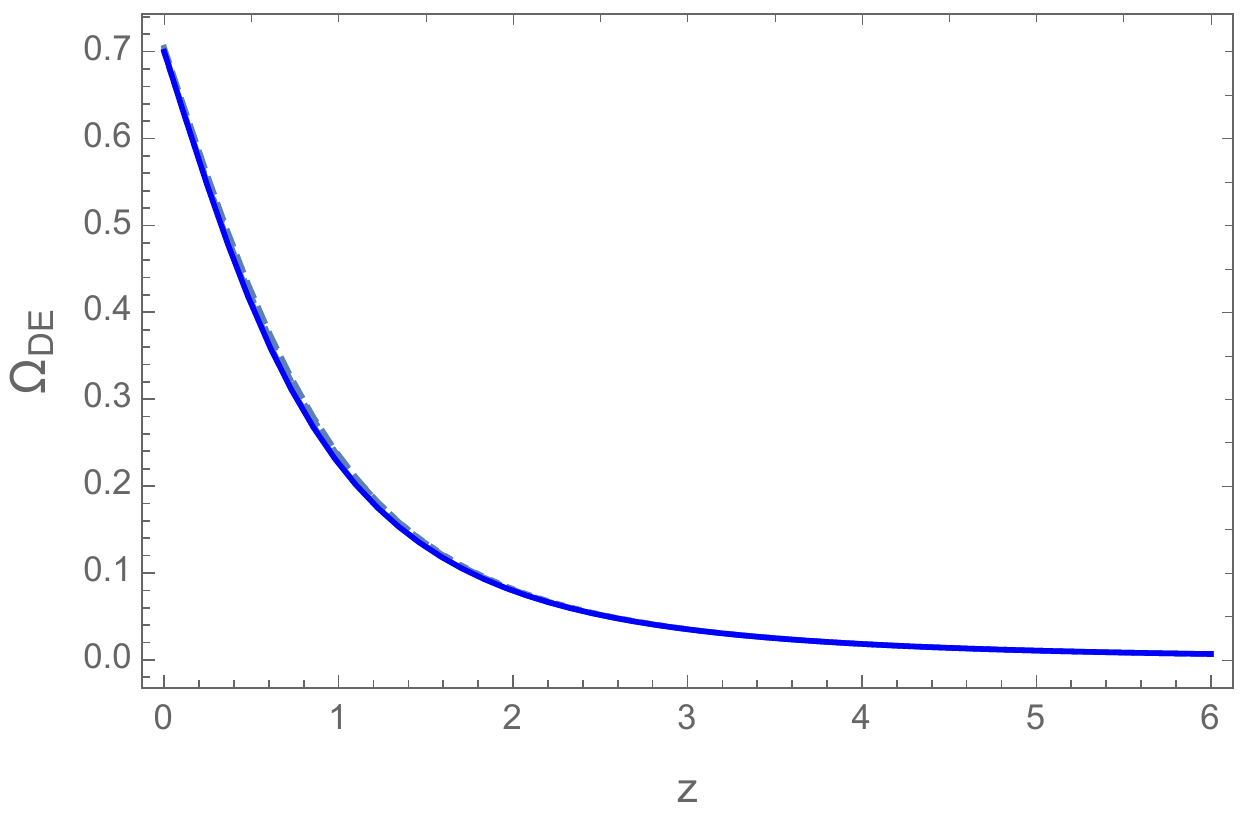}
\includegraphics[scale=0.57]{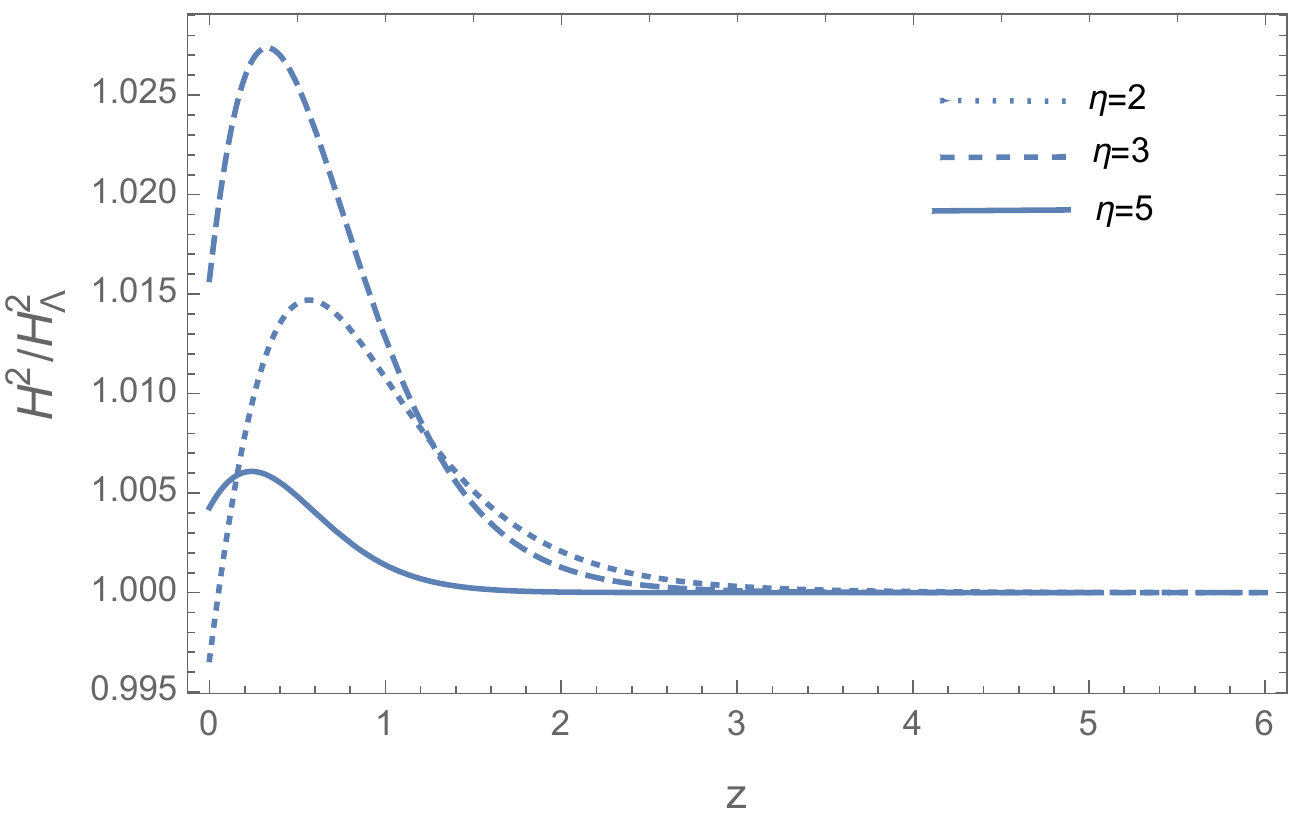}
\caption{The evolution of $y_H, w_{DE}, \Omega_{DE}$, and $H^2/H_{\Lambda}^2$. The following initial redshifts have been used:  $z_i=10$, for $\eta=2$ and $\lambda_1=15$ (dotted), $z_i=7.67$ for $\eta=3$ and $\lambda_1=500$ (dashed) )and $z_i=35.6$ for $\eta=5$ and $\lambda_1=5\times 10^4$ (solid). The horizontal line in $w_{DE}$ corresponds to the cosmological constant.} 
\label{fig2}
\end{center}
\end{figure}  
\noindent The background evolution of DE density $\Omega_{DE}$ is indistinguishable from that of $\Lambda$CDM model (blue curve) in Figs. 1 and 2. A very small difference can be appreciated in the ratio $H^2/H_{\Lambda}^2$. For $\lambda_1=1$ the closeness to $\Lambda$CDM is accentuated as $\eta$ increases. 
Note the absence of DE oscillations in $y_H$ but they are present in $w_{DE}$ since $y'_H$ in (\ref{back10}) involves higher derivatives of $H$. The amplitude of these DE oscillations can be significantly reduced with the adequate choice of $\lambda_1$. The above numerical results show that the background evolution of the model (\ref{model2}) cannot bring appreciable differences with the standard $\Lambda$CDM model, which is a typical feature of $f(R)$ models. Characteristic patterns must be looked for in the evolution of matter perturbations.
\section{Constraints from Matter Density Perturbations}
The evolution of matter density perturbations  \cite{stsujikawa1, stsujikawa2, stsujikawa0} lead to observational signatures of $f(R)$ models of dark energy that distinguish them from the $\Lambda$CDM model. For the wave number $k$ deep inside the Hubble radius ($k>>aH$) the equation of matter perturbations can be reduced to the following
\be\label{matter-pert}
\ddot{\delta}_m+2H\dot{\delta}_m-4\pi G_{eff}\rho_m\delta_m=0
\ee
where $\delta_m=\delta\rho_m/\rho_m$ and $G_{eff}$ is the effective gravitational coupling that encodes the effect of $f(R)$ and is defined by 
\be\label{eff-newton}
G_{eff}=\frac{G}{f_{,R}}\left[\frac{1+4\frac{k^2f_{,RR}}{a^2f_{,R}}}{1+3\frac{k^2f_{,RR}}{a^2f_{,R}}}\right],
\ee
which can be approximated, in the matter dominated era where $f_{,R}\simeq 1$ for viable models, as
\be\label{eff-newton1}
G_{eff}\simeq G\left[\frac{1+\frac{4k^2 m}{a^2R}}{1+\frac{3k^2m}{a^2R}}\right]\simeq G\left[\frac{1+\frac{4k^2}{3a^2M^2}}{1+\frac{k^2}{a^2M2}}\right]
\ee
where (\ref{effmass1}) was used. \\
In the scale of validity of linear regime ($10^{-2}Mpc^{-1}\lesssim k\lesssim 0.15 Mpc^{-1}$) \cite{linder, percival, huterer, tsujikawa3}, some $f(R)$ may behave in such a way that can be distinguished from $\Lambda$CDM and this effect could be sensitive to near future observations. 
The transition from GR regime to scalar tensor regime may be established when the effective mass scale $M$ is comparable to the inverse of the scale length $\ell=a/k$, i.e.  
\be\label{growth}
\ell^2 M^2\approx 1\;\; \Rightarrow\;\; M^2\simeq \frac{k^2}{a^2},\;\;\; \Rightarrow m\approx \left(\frac{aH}{k}\right)^2,
\ee
where the last relation follows from (\ref{effmass1}). According to this, in order that a region of size $\ell$ is not affected by modifications of gravity it must be $\ell M>>1$, which applied to cosmic scales means that effects of modified gravity can appear at scales smaller than $M^{-1}$. If we consider a galaxy cluster with a size $\ell$ of the order of 10Mpc, then the following relations apply
\be\label{growth1}
\ell\sim\frac{a}{k}\sim \frac{1}{H}=10 Mpc\approx 3.26\times 10^7 ly.
\ee
If $\ell_0$ is the size of the universe, then 
\be\label{growth2i}
\ell_0\sim\frac{a_0}{k_0}\sim \frac{1}{H_0}\approx 1.38\times 10^{10} ly,
\ee
and
\be\label{growth3i}
\frac{k}{k_0}=\frac{\ell_0}{\ell}\approx 4.2\times 10^2\Rightarrow k\approx 420 k_0\approx 420 a_0 H_0
\ee
this wave number is near the upper limit of the scale relevant to the galaxy power spectrum as cited above, which is between the validity of the linear regime \cite{percival}.
Thus, if the transition to scalar-tensor regime occurred in the representative current epoch, then $m(z\sim 0)$ should satisfy the condition
\be\label{growth4i}
m(z\approx 0)\gtrsim (420)^{-2}\approx 5.6\times 10^{-6},
\ee
in order to find deviations from GR at scales $k\sim 0.1 Mpc^{-1}$. Numerical results (see table I) show that $m\gtrsim 10^{-6}$ can be achieved at current times, while still satisfying local gravity restrictions, for $\eta>1$.
If the transition to scalar-tensor regime occurs during deep matter era \cite{tsujikawa1, tsujikawa3} then the transition redshift $z_k$ can be estimated using the approximation valid during matter dominance
\be\label{matter1}
H^2\simeq H_0^2\Omega_{m0}\left(1+z\right)^3,\;\;\; R\simeq 3H^2,
\ee
and the approximation for $m$ from (\ref{my1}), valid for $R>>\mu^2$
\be\label{mmatter}
m\approx \frac{1}{2}\lambda_1\eta\left(\eta+1\right)y_{ds}\left(\frac{\mu^2}{R}\right)^{\eta+1},
\ee
which lead, using (\ref{growth}), to
\be\label{ztransition}
z_k=\left[\left(\frac{k}{a_0H_0}\right)^2\frac{\lambda_1\eta(\eta+1)y_{ds}}{2\Omega_{m0}3^{\eta+1}}\right]^{\frac{1}{3\eta+4}}-1,
\ee
where we used $\mu^2=\Omega_{m0}H_0^2$. In table II we show some results for the wave number $k=300 a_0H_0$ calculated for $\Omega_m=0.3$.\\
\noindent One can also appreciate the effect of the scalar-tensor regime on the matter power spectrum compared to the effect of the $\Lambda$CDM model. First we note that in the scalar-tensor regime the effective gravitational coupling becomes $G_{eff}\simeq 4G/(3f_{,R})\simeq 4G/3$ and the evolution of matter density perturbations behaves as $\delta_m\propto t^{(\sqrt{33}-1)/6}$ (valid for $t_k<t<t_{\Lambda}$ where at $t_{\Lambda}$ the transition to accelerated expansion occurs. i.e $\ddot{a}=0$), while in the $\Lambda$CDM model $\delta_m\propto t^{2/3}$  \cite{stsujikawa1, stsujikawa2, stsujikawa0}. Taking into account that during matter dominance $a=(1+z)^{-1}\propto t^{2/3}$, then from (\ref{ztransition}) follows that the time $t_k$ has the scale dependence $t_k\propto k^{-\frac{3}{3\eta+4}}$. Then this scale dependence of the scalar-tensor phase induces modification of the matter power spectrum compared to the $\Lambda$CDM model, which at time $t=t_{\Lambda}$ leads to  \cite{stsujikawa1, stsujikawa2, stsujikawa0}
\be\label{spectra}
\frac{P_{\delta_m}(t_{\Lambda})}{P^{\Lambda}_{\delta_m}(t_{\Lambda})}=\frac{|\delta_m|^2}{|\delta_m^{\Lambda}|^2}\propto k^{\frac{\sqrt{33}-5}{3\eta+4}}
\ee
This equation gives rise to a difference between the spectral indices of the matter power spectrum and of the CMB spectrum on scales relevant to the galaxy power spectrum  ($10^{-2}Mpc^{-1}\lesssim k\lesssim 0.15 Mpc^{-1}$), given by
\be\label{spectra1}
\Delta n(t_{\Lambda})=\frac{\sqrt{33}-5}{3\eta+4}.
\ee
Some cases are shown in table II.
\begin{center}
\begin{tabular}{lccccc}\hline\hline
 $\eta$  \hspace{0.6cm} & $\lambda_1$  \hspace{0.6cm}  &$z_k$ ($\frac{k}{a_0H_0}\approx 300$)  \hspace{0.6cm} & $z_k$ ($\frac{k}{a_0H_0}\approx 600$)\hspace{0.3cm} & $\Delta n(t_{\Lambda})$ \\ 
\hline
$1$ & $1$ & $6.13$ & $7.7$ & $0.106$  \\ 
$2$& $1$  &$2.95$ &$3.54$ &$0.074$\\
$2$& $15$  &$4.18$ &$4.95$ &$0.074$\\
 $3$  & $1$ & $1.79$ & $2.1$&  $0.057$\\ 
 $3$ & $5\times 10^2$ & $3.5$ & $4$ & $0.057$\\  
 $4$ & $5\times 10^3$ & $2.78$ & $3.12$ & $0.046$\\  
$5$  & $5\times 10^4$ & $2.33$  & $2.59$ & $0.039$\\ 
 \hline
\end{tabular}
\end{center}
\begin{center}
{\bf Table II}
\end{center}
\noindent {\it The redshift transition for the modes $k=300 a_0H_0\simeq 0.1h Mpc^{-1}$ and $k=600 a_0H_0\simeq 0.2h Mpc^{-1}$. These values indicate that the transition to scalar-tensor regime occurred during matter dominated era. $z_k$ depends on $\eta$ and $k$, being larger for larger $k$, for a given mode. The bound $\Delta n(t_{\Lambda})<0.05$ is satisfied starting from $\eta>3$.}\\

\noindent Using the criterion $\Delta n(t_{\Lambda})<0.05$ \cite{astarobinsky, tsujikawa1}, according to results in table II we find the bound $\eta>3$. \\

\noindent {\bf The Growth of Matter Perturbations.}\\

\noindent  Apart from the cosmic expansion history, the growth of large scale structure in the universe provides an important test which can reveal a deviation from the $\Lambda$CDM model especially at late times. 
The Eq. (\ref{matter-pert}) for the fractional matter density perturbation $\delta_m$ can be written in terms of the $e$-fold variable $N=\ln a$ as follows
\be\label{growth}
\frac{df(a)}{dN}+f(a)^2+\frac{1}{2}\left(1-\frac{d\ln \Omega_m(a)}{dN}\right)f(a)=\frac{3}{2}\frac{G_{eff}}{G}\Omega_m(a)
\ee
where 
\be\label{growth1i}
f(a)=\frac{d\ln\delta_m}{dN}
\ee
and $\Omega_m(a)$ is given by
\be\label{growth2}
\Omega_m(a)=\frac{\kappa^2\rho_m}{3H^2}=\frac{\Omega_{m0}a^{-3}}{H^2/H_0^2},\;\;\; \Omega_{m0}=\frac{\kappa^3\rho_{m0}}{3H_0^2},
\ee
where the second equality takes place for dust matter, neglecting the radiation.
The function $f$ can be written in the form
\be\label{growth3}
f(a)=\Omega_m(a)^{\gamma(a)},
\ee
where $\gamma$ defined by
\be\label{growth4}
\gamma(a)=\frac{\ln f(a)}{\ln \Omega(a)}
\ee
is the growth index of matter perturbations \cite{peebles,wangstein,linder}. In order to integrate the eq. (\ref{growth}) in the matter dominated epoch we use the fact that in the high redshift region the model (\ref{model2}) is close to the $\Lambda$CDM model, and therefore we can assume that the background expansion is well approximated by the $\Lambda$CDM model. The following approximation takes place for $R>>\mu^2$
\be\label{growth5}
\frac{f_{,RR}}{f_{,R}}=\frac{m}{R}\approx \frac{\lambda\lambda_1\eta(\eta+1)}{R}\left(\frac{\mu^2}{R}\right)^{\eta+1},
\ee
which is used in $G_{eff}$ given in (\ref{eff-newton1}). In Figs. 3-5 we show the evolution of the growth function $f$ and the growth index $\gamma$  for some cases with $\lambda_1>1$ (see Fig. 2), for the modes $k/(a_0H_0)=30, 100, 300, 600$. 
\begin{figure}
\begin{center}
\includegraphics[scale=0.57]{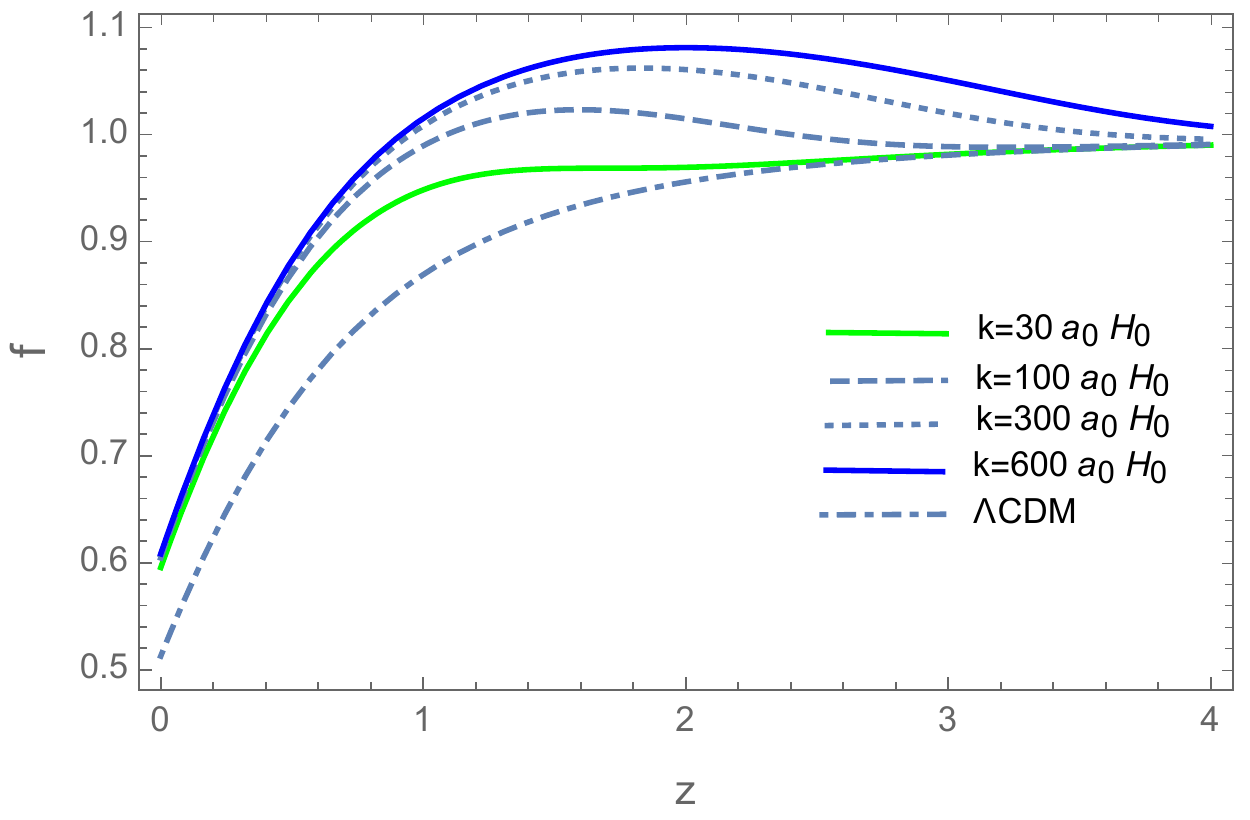}
\includegraphics[scale=0.57]{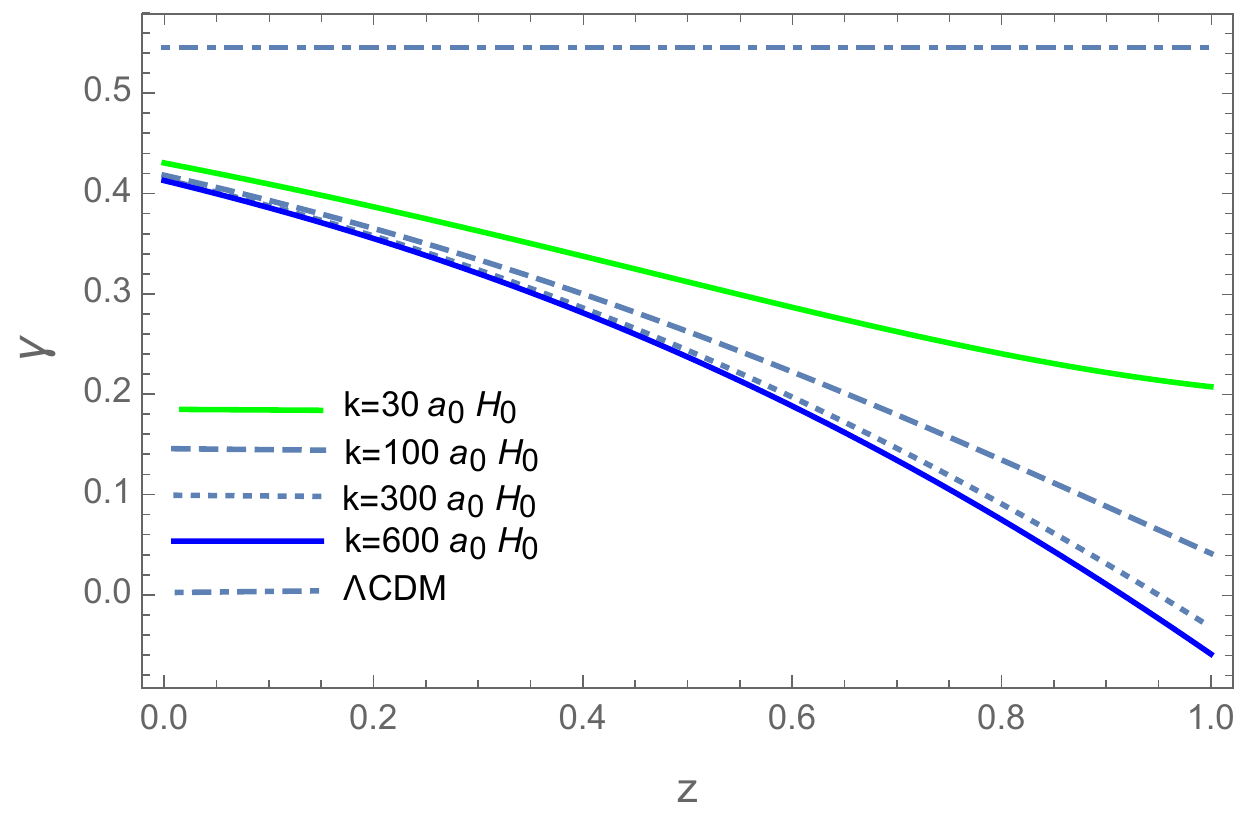}
\caption{The evolution of the growth rate  $f$ and the growth index $\gamma$ in the model (\ref{model2}) for four different values of $k$. The curves correspond to $\eta=3$ and $\lambda_1=5\times 10^2$, with $\mu^2=\Omega_{m0}H_0^2$ and $\Omega_m=0.3$. The dispersion in $k$ practically disappears at low redshifts. The transition redshifts for the different modes are: $k=30 a_0H_0,\; z_k=2.16$,  $k=100 a_0H_0,\; z_k=2.8$,  $k=300 a_0H_0,\; z_k=3.5$ and $k=600 a_0H_0,\; z_k=4$. The larger the scale, the later the growth rate enters the transition regime.}
\label{fig3}
\end{center}
\end{figure}

\begin{figure}
\begin{center}
\includegraphics[scale=0.59]{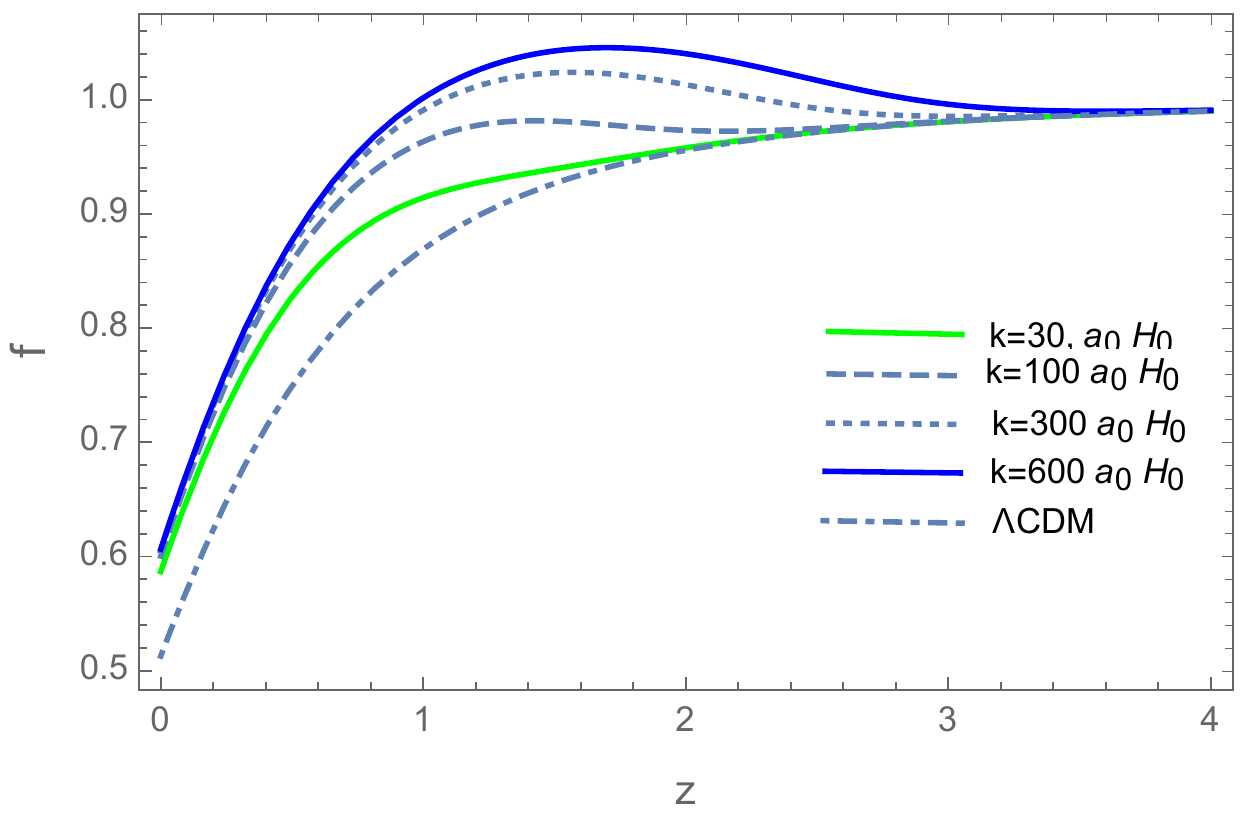}
\includegraphics[scale=0.59]{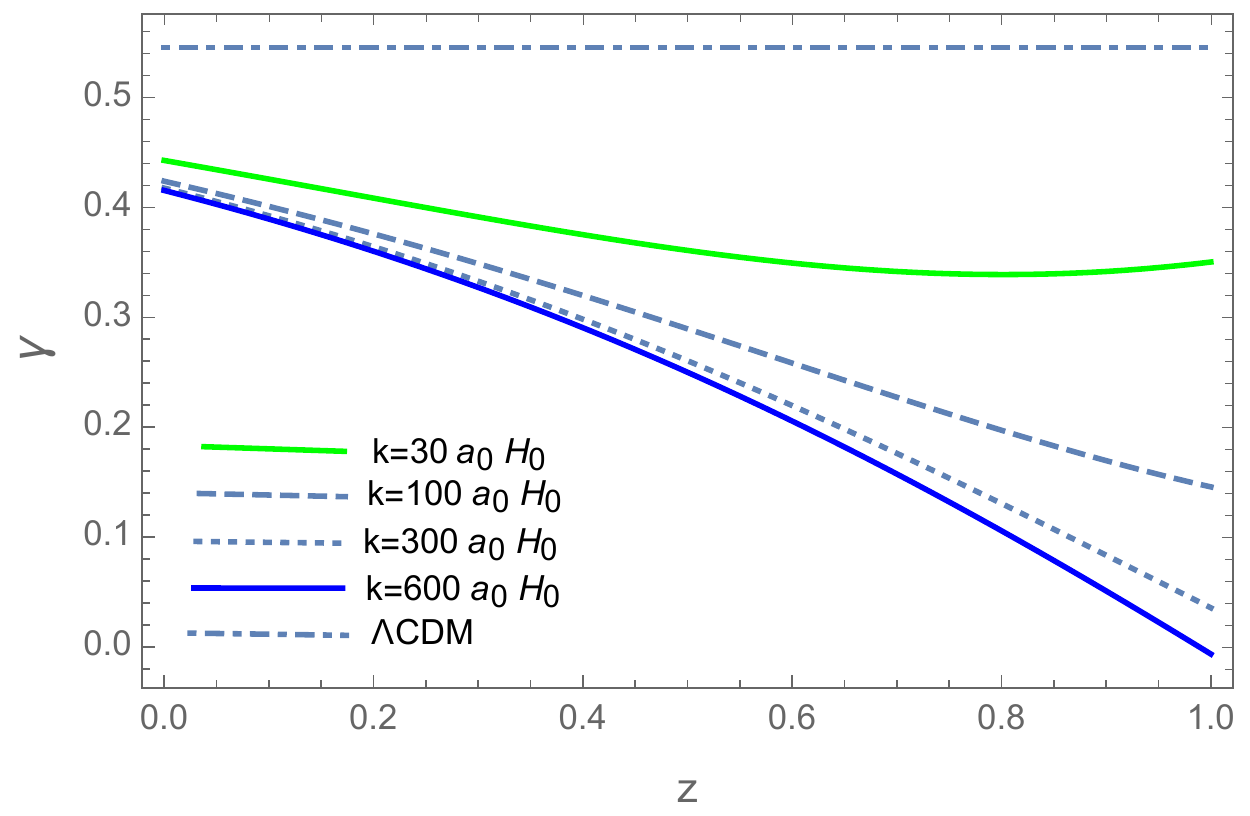}
\caption{The evolution of the growth index $\gamma$ for the cases $\eta=4,\; \lambda_1=5\times 10^{3}$. The near scale invariance is kept at low redshifts.
The transition redshifts for the different modes are: $k=30 a_0H_0,\; z_k=1.83$,  $k=100 a_0H_0,\; z_k=2.29$,  $k=300 a_0H_0,\; z_k=2.78$ and $k=600 a_0H_0,\; z_k=3.12$. All scales enter the transition regime later than the corresponding ones in Fig. 3.}
\label{fig4}
\end{center}
\end{figure}

 \begin{figure}
\begin{center}
\includegraphics[scale=0.59]{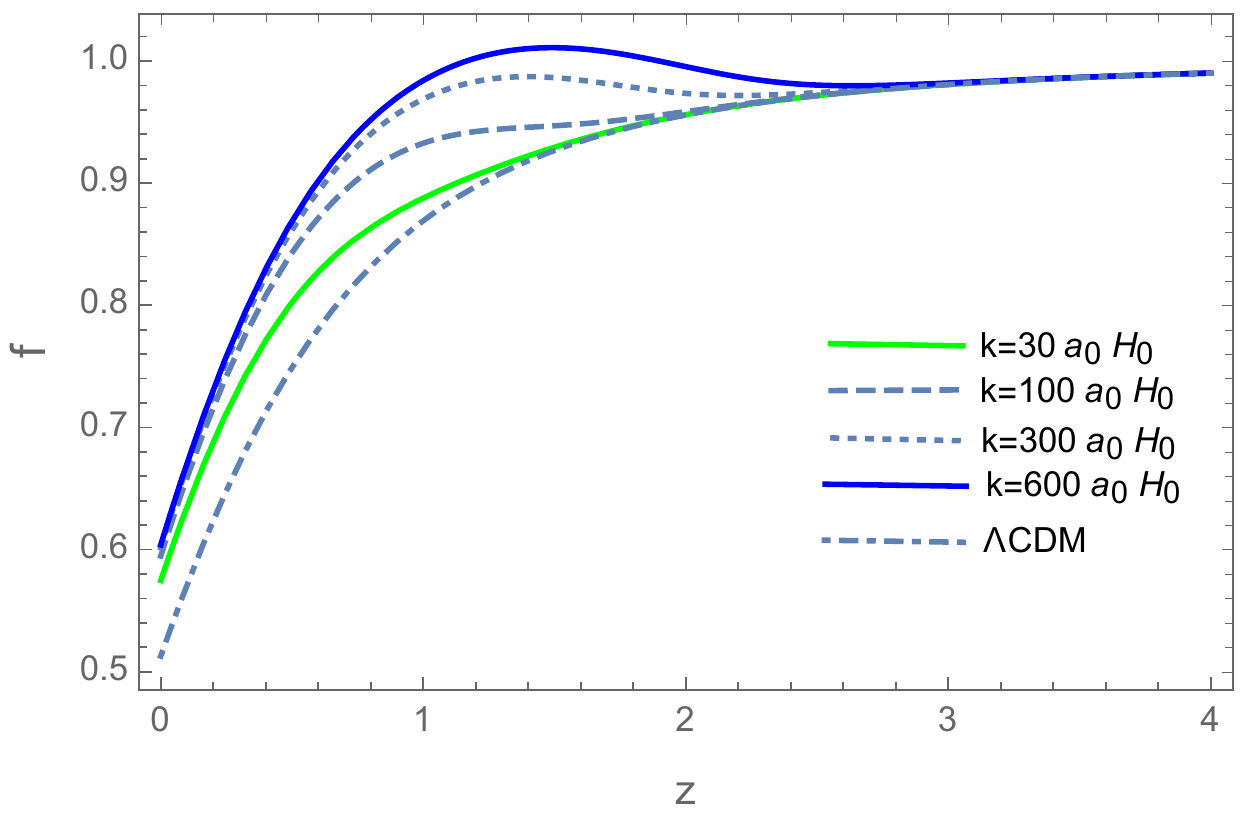}
\includegraphics[scale=0.59]{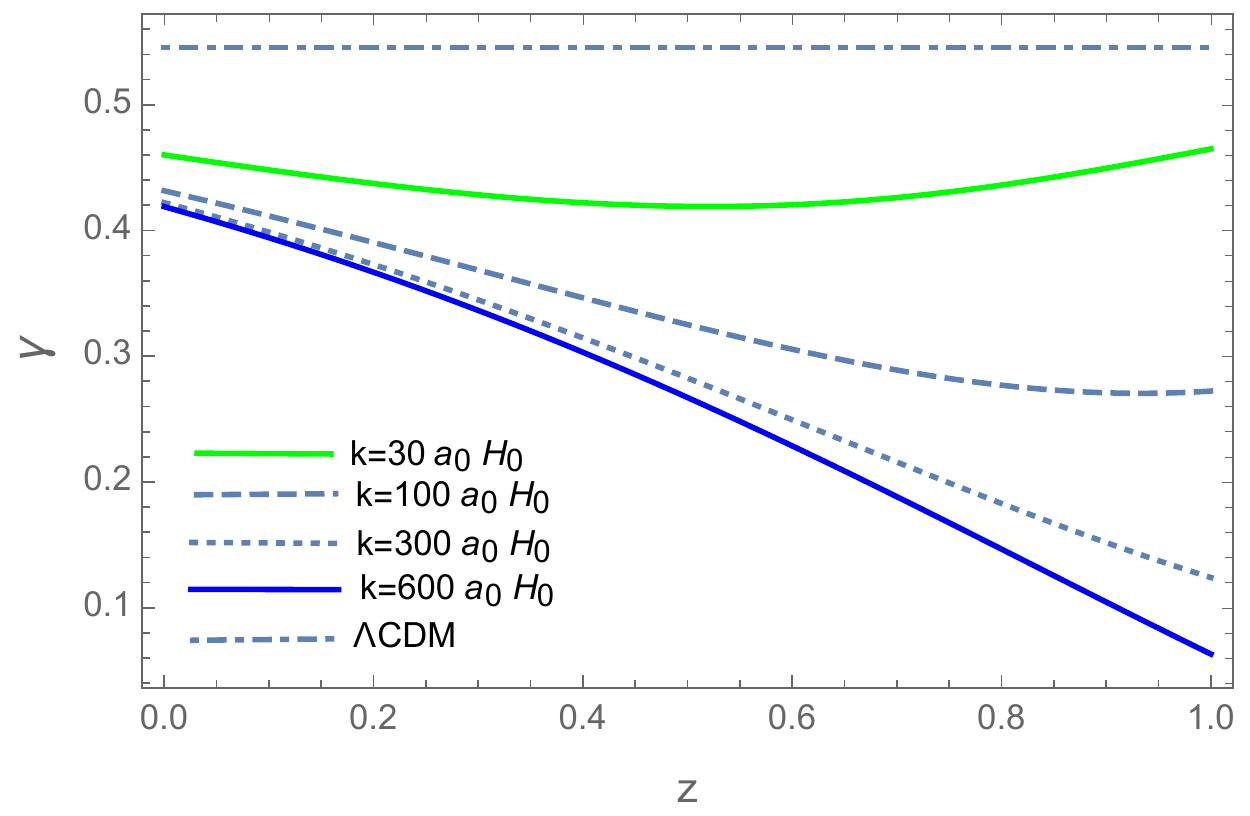}
\caption{The evolution of $f$ and $\gamma$ for the model (\ref{model2}) assuming $\eta=5,\; \lambda_1=5\times 10^4$. A small dispersion at low redshift is more noticeable in $\gamma$ for the larger scale. The transition redshift for the different modes are: $k=30 a_0H_0,\; z_k=1.62$,  $k=100 a_0H_0,\; z_k=1.97$,  $k=300 a_0H_0,\; z_k=2.33$ and $k=600 a_0H_0,\; z_k=2.59$. All transition redshifts move to lower values (later times) compared to the cases of Figs. 3 and 4.}
\label{fig5}
\end{center}
\end{figure}

 \begin{figure}
\begin{center}
\includegraphics[scale=0.59]{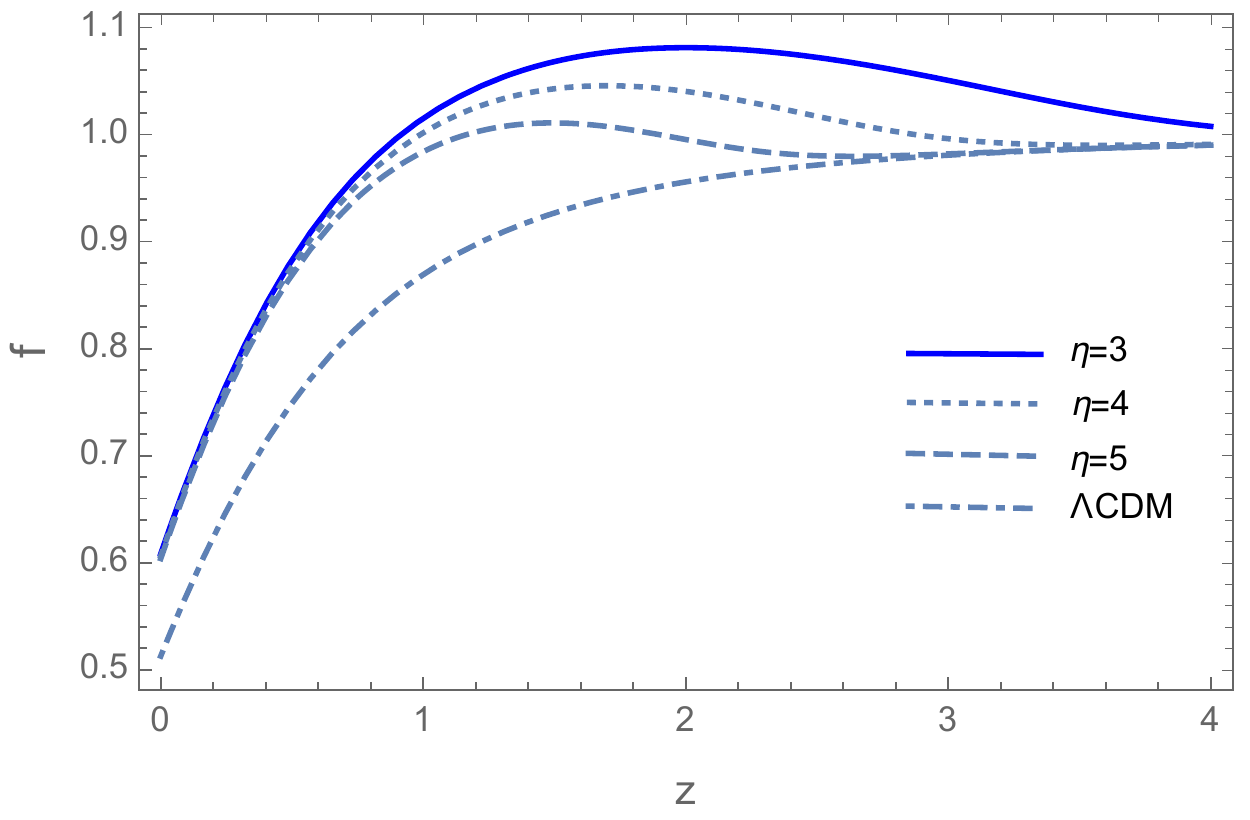}
\includegraphics[scale=0.59]{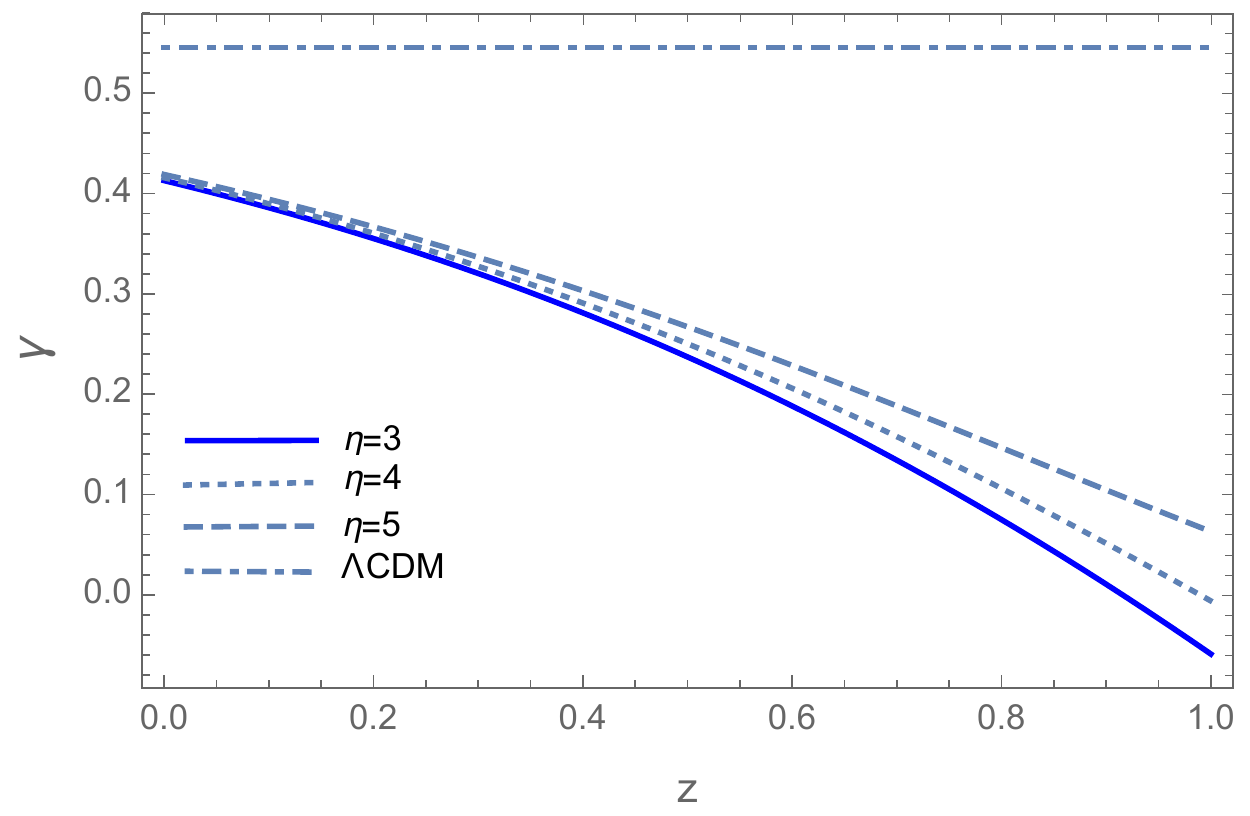}
\caption{Comparison of the evolution of $f$ and $\gamma$ for different values of $\eta$. The curves correspond to the mode $k=600 a_0H_0$, assuming $\lambda_1=5\times 10^2, 5\times 10^3, 5\times 10^4 $ for $\eta=3,4,5$ respectively. The maximum is higher for lower values of $\eta$ and the bigger $\eta$ the later the maximum is reached. Note also the absence of $\eta$-dependence at low redshifts.}
\label{fig6}
\end{center}
\end{figure}
\noindent Comparing the curves in Figs. 3-5 it can be seen that as the curves get closer to $\Lambda$CDM, a slight increase in dispersion is observed at low redshifts, which is more notable for the larger scale. A quantity that can be used to the analysis is 
\be\label{q}
q=\frac{\lambda_1}{y_{ds}^{\eta}}. 
\ee
From the expression for $\lambda$ given in (\ref{lambda-ds1}) it follows that the approximation $\lambda\approx y_{ds}/2$ takes place only in the case $q<<1$. For the numerical examples of Figs. 3-5 we have $q=0.023$ for $\eta=3$, $q=0.008$ for $\eta=4$ and $q=0.003$ for $\eta=5$. In fact if one takes the same $q<<1$ for all cases, varying only eta, we can observe curves like those depicted in Figs. 3-5 with the following characteristics, focusing on separate modes: the width of the crest becomes shorter as $\eta$ grows and the maximum of $f$ decreases as $\eta$ grows. The transition redshift $z_k$ moves to lower values as $\eta$ increases. These features can be seen in Fig. 6. Note also that $\lambda_1$ cannot take arbitrary large values since it is restricted by the inequality (\ref{ds-stability}). 
An appropriate approximate relation between the current value of the deviation parameter, i.e. $m(y_0)$ and $q$ can be deduced if we assume that $m(y_0)\sim m(y_ds)$. From (\ref{my1}) follows
\be\label{m-ds}
m(y_ds)=\frac{\eta q\left(1+\eta-\eta q\right)}{2-2\eta q}\approx \frac{1}{2}\eta(\eta+1)q\sim m(y_0),
\ee
where the approximation takes place if we assume $q<<1$. Taking into account the bound (\ref{growth4i}) it is found the bound on $q$
\be\label{q1}
q\gtrsim \frac{10^{-5}}{\eta(\eta+1)},
\ee
then, choosing values for $q$ of the order of $10^{-2} - 10^{-4}$ with $\eta$ such that the bound (\ref{growth4i}) is satisfied leads to characteristic signatures of the model in the evolution of matter perturbations. \\
\noindent In all cases considered for $\lambda_1>1$ the domain of $\gamma(z)$ is far from $\gamma_0$ for $\Lambda$CDM at low redshifts and the dispersion of $\gamma(z\approx 0)$ is very small. This suggests that all scales analyzed in Figs. $3 - 5$ have reached the asymptotic regime $k>>aM$ at current epoch (the largest scale still shows some dispersion). It is also remarkable that the value $\gamma(z\approx 0)$ is independent of the model parameter $\eta$ for a given mode as can be sen from Fig. 6, which also takes place approximately independently of the mode by looking at Figs. 3-5. Therefore the value $\gamma(z\approx 0)$ is essentially independent of the model parameter $\eta$. The facts that the current value of the growth index ($\gamma(z\approx 0)$) is scale independent, and that is significantly lower than $\gamma_0$ for $\Lambda$CDM, is a characteristic signature of the model (\ref{model2}) as an $f(R)$ model.\\
Numerical analysis also shows that the models with $\lambda_1=1$ and $\eta\ge3$ (for which Solar system constraints are fulfilled, i.e. $m(y_s)<10^{-23}$)
do not exhibit visible patterns in the evolution of the growth of matter perturbations compared to $\Lambda$CDM, which is accentuated as $\eta$ increases. The main difference appears in the growth index, which turns to the opposite direction, crossing the $\gamma_0$ line and becoming a bit larger than $\gamma_0$. In Fig. 8 we illustrate the behavior of $\gamma(z)$ for $\eta= 4,5,6$ and $\eta\ge7$.\\


\begin{figure}
\begin{center}
\includegraphics[scale=0.57]{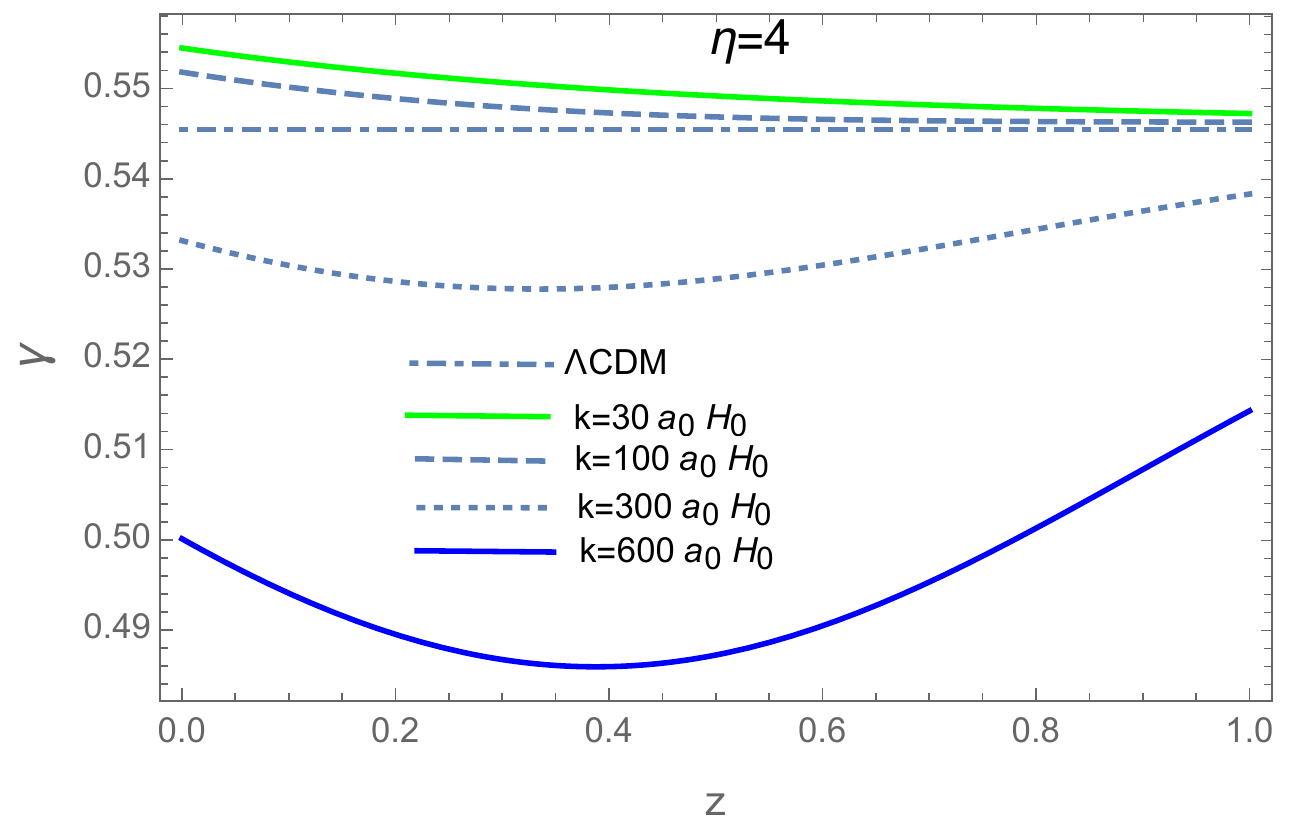}
\includegraphics[scale=0.57]{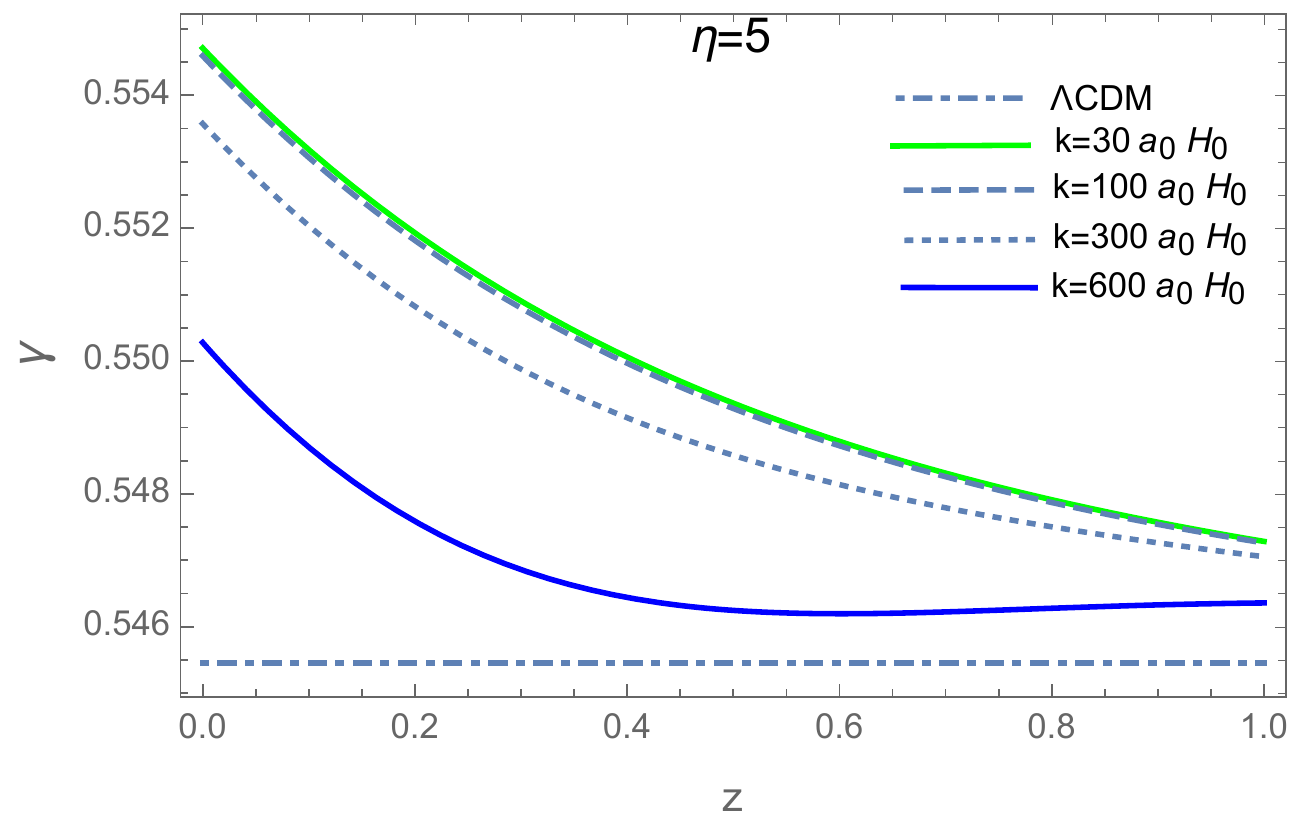}
\includegraphics[scale=0.57]{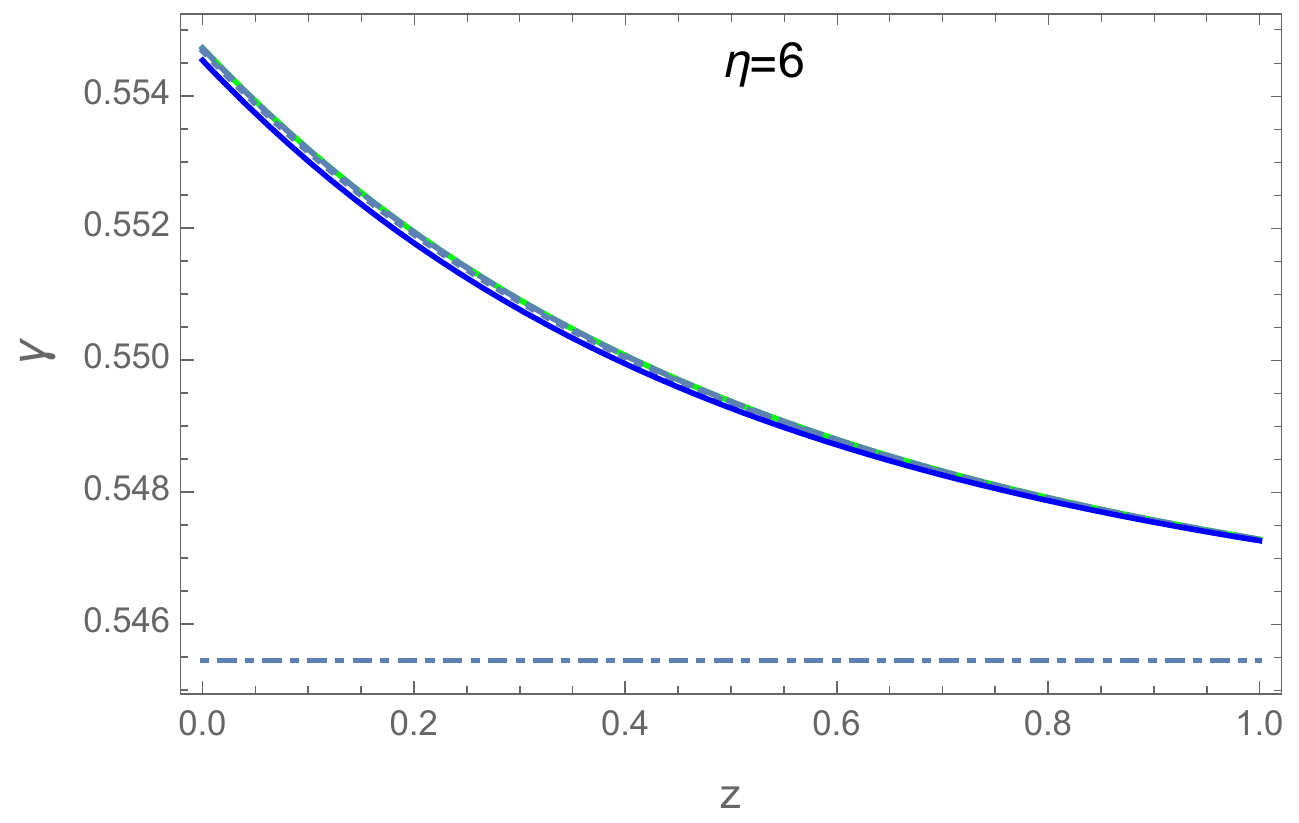}
\includegraphics[scale=0.57]{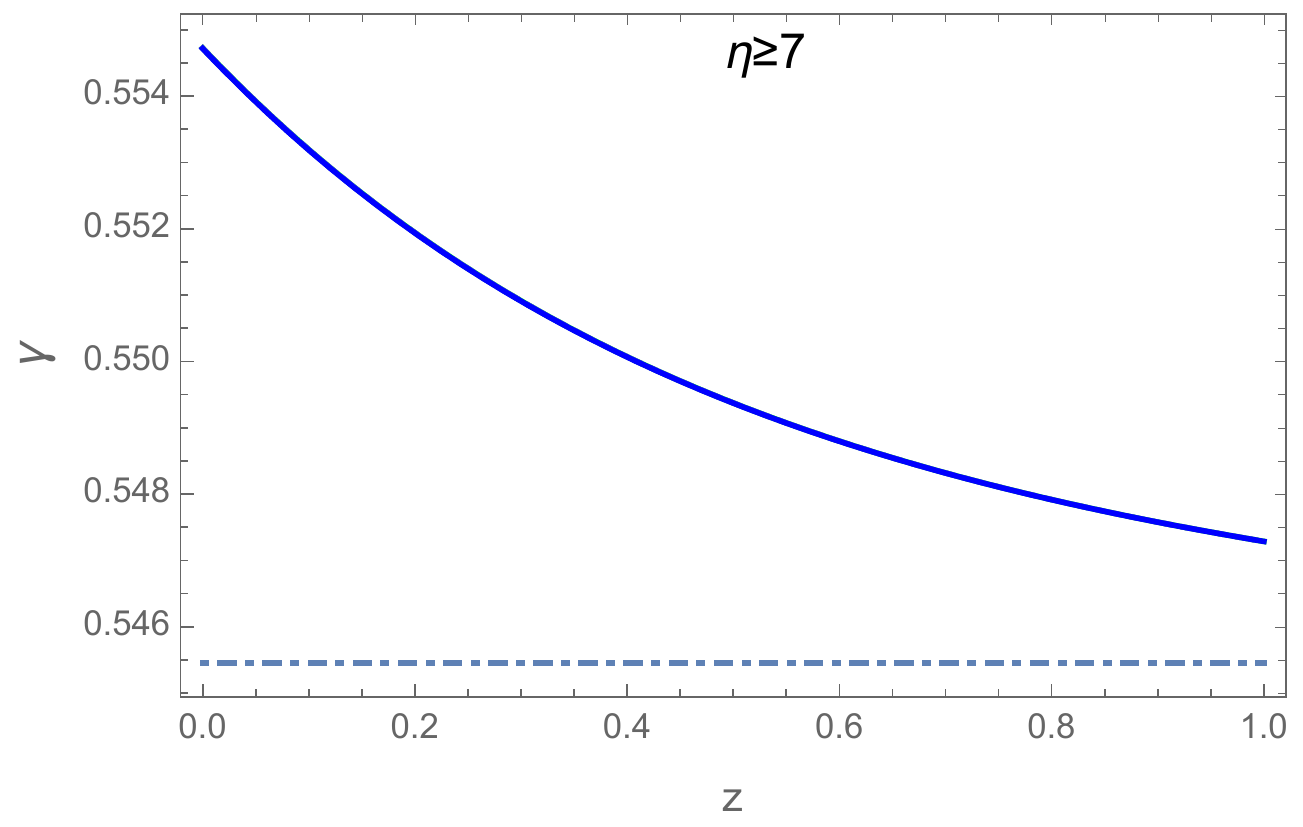}
\caption{The evolution of the growth index $\gamma$ in the model (\ref{model2}) for  $\lambda_1=1$ and four different values of $k$. The curves on the left correspond to and $\eta=4$ and the curves on the right to $\eta=5$. The dispersion of $\gamma$ is evident for $\eta=4, 5$, decreasing for $\eta>5$ until it disappears for $\eta\ge 7.$ Then, for $\eta\ge 7$ the model becomes scale invariant and the only difference with $\Lambda$CDM is in the behavior of $\gamma(z)$.}
\label{fig7}
\end{center}
\end{figure}


\newpage
\noindent {\bf f$\sigma_8$ Evolution of the Model}\\

\noindent The weighted growth rate, expressed as $f\sigma_8(a)$, has become an important cosmological observable because this product is independent of the bias factor between the observed galaxy spectrum and the underlying matter power spectrum \cite{percival1}. $\sigma_8(a)$ is the matter power spectrum normalization on scales of $8h^{-1}Mpc$.
It is well known that the $\Lambda$CDM model predicts values for $\sigma_8$, which lead to an exceeding structure formation power, entering in tension with LSS observations \cite{planck2015}. The weighted growth rate can be expressed as
\be\label{sigma8}
f(a)\sigma_8(a)=\frac{\sigma_8}{\delta(1)}a\delta'(a)=-\frac{\sigma_8}{\sigma(0)}(1+z)\frac{d\delta (z)}{dz}
\ee
where $\sigma_8(a)=\sigma_8 \delta(a)/\delta(1)$ is the r.m.s. fluctuation of density perturbations on scale  $8h^{-1}Mpc$ and $\sigma_8$ is its current value.
In redshift variable the Eq. (\ref{matter-pert}) reads
\be\label{deltam}
(1+z)^2\delta''(z)+(1+z)\left[\frac{3}{2}\Omega_m(z)-1\right]\delta'(z)-\frac{3}{2}\frac{G_{eff}}{G}\Omega_m(z)\delta(z)=0
\ee
where we have assumed, according to the results illustrated in Figs. 1 and 2, the $\Lambda$CDM Hubble parameter for the background evolution. 
Numerical solution of this equation together with Eq. (\ref{sigma8}), with initial conditions in the deep matter era $\delta(z_i)\sim 1/z_i$ and $\delta'(z_i)\approx 0$ (taking $z_i\sim 50-100$), gives the theoretical prediction of the model (\ref{model2}) (using (\ref{mmatter})) for $f\sigma_8(z)$. In Figs. 7 and 8 we plot the evolution of the weighted growth rate $f\sigma_8(z)$ contrasted with the full data set of 63 $f\sigma_8$ measurements from various surveys \cite{percival1, davis, hudson, turnbull, samushia, blake, beutler, rita, torre, chia, blake1, ariel, howlett, feix, okumura, chia1, alam,beutler1, wilson, marin, hawken, huterer1, torre1, pezzotta, feix1, howlett1, mohammad, wang, shi, marin2, hou, bo}. For the theoretical curves predicted by the model (\ref{model2}) we have assumed $\sigma_8=0.82$ consistent with Planck15/$\Lambda$CDM data \cite{planck2015}. Considering the model in the context of all available data set allows to find out the ability of the model to fit the observational data, and therefore, to increase or reduce the tension. 
 \begin{figure}
\begin{center}
\includegraphics[scale=0.6]{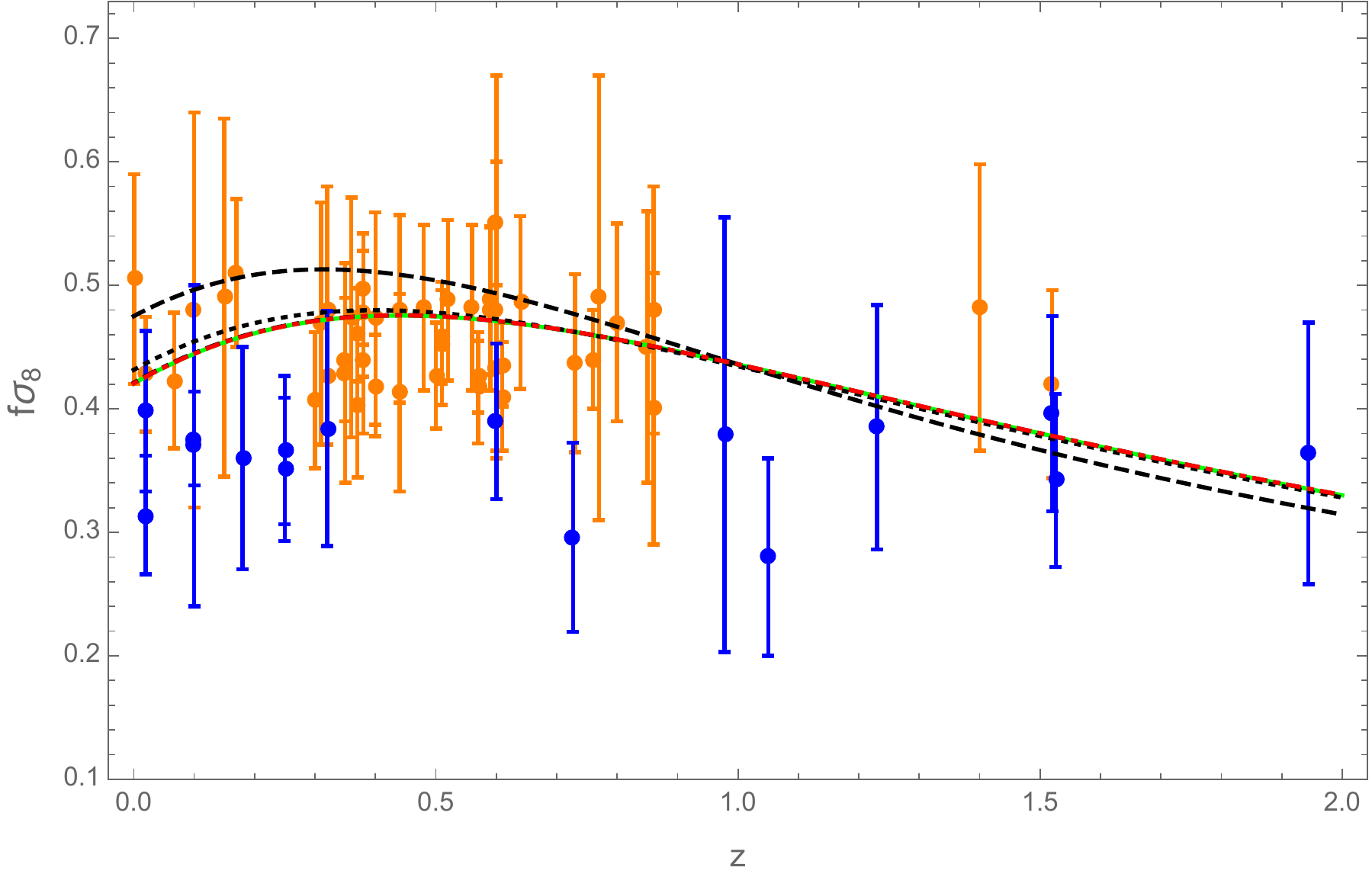}
\includegraphics[scale=0.6]{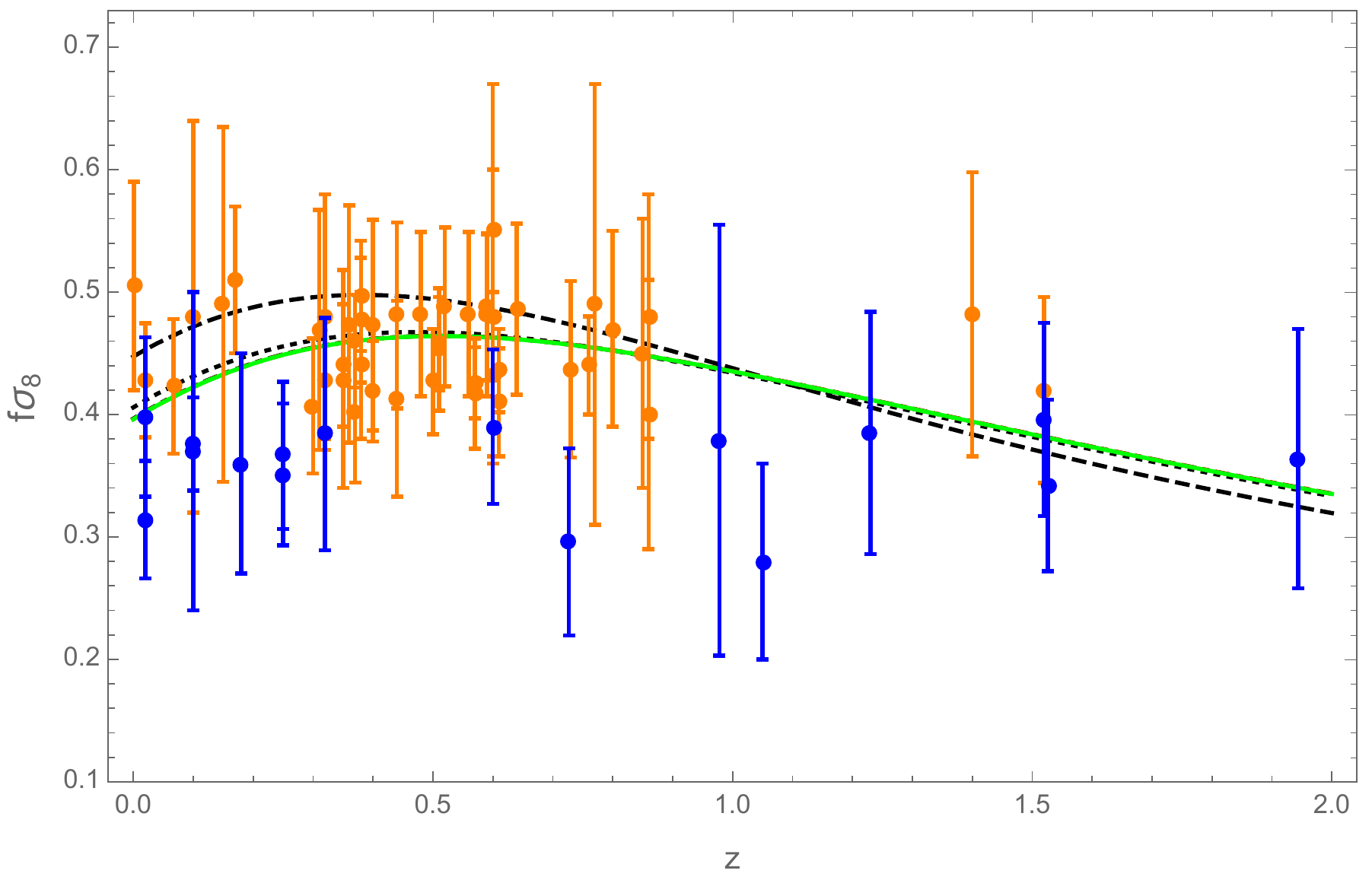}
\caption{The evolution of $f\sigma_8$ for the model (\ref{model2}) taking $\lambda_1=1$ and $\eta=3$(dashed), $4$(dotted), $5$(dot-dashed), $7$(red), contrasted with the observed values of $f\sigma_8$ from the cited surveys. The orange error bars correspond to main values $f\sigma_8>0.4$ and the blue ones to $f\sigma_8<0.4$, which highlights the tension with the low redshift data mostly for $f\sigma_8<0.4$. The curves correspond to the mode $k=300 a_0H_0$, assuming $\Omega_{m0}=0.3$ and initial conditions for Eq. (\ref{deltam}) at $z_i=50$. The green curve corresponds to $\Lambda$CDM. The lower curves correspond to $\Omega_{m0}=0.27$, which lowers $f\sigma_8(z)$ a bit, suggesting that the observed values of $f\sigma_8$ favor a darker universe. It is clear that the curves do not improve the $\sigma_8$-tension and rather align with $\Lambda$CDM for $\eta\ge 4$.}
\label{fig9}
\end{center}
\end{figure}

 \begin{figure}
\begin{center}
\includegraphics[scale=0.6]{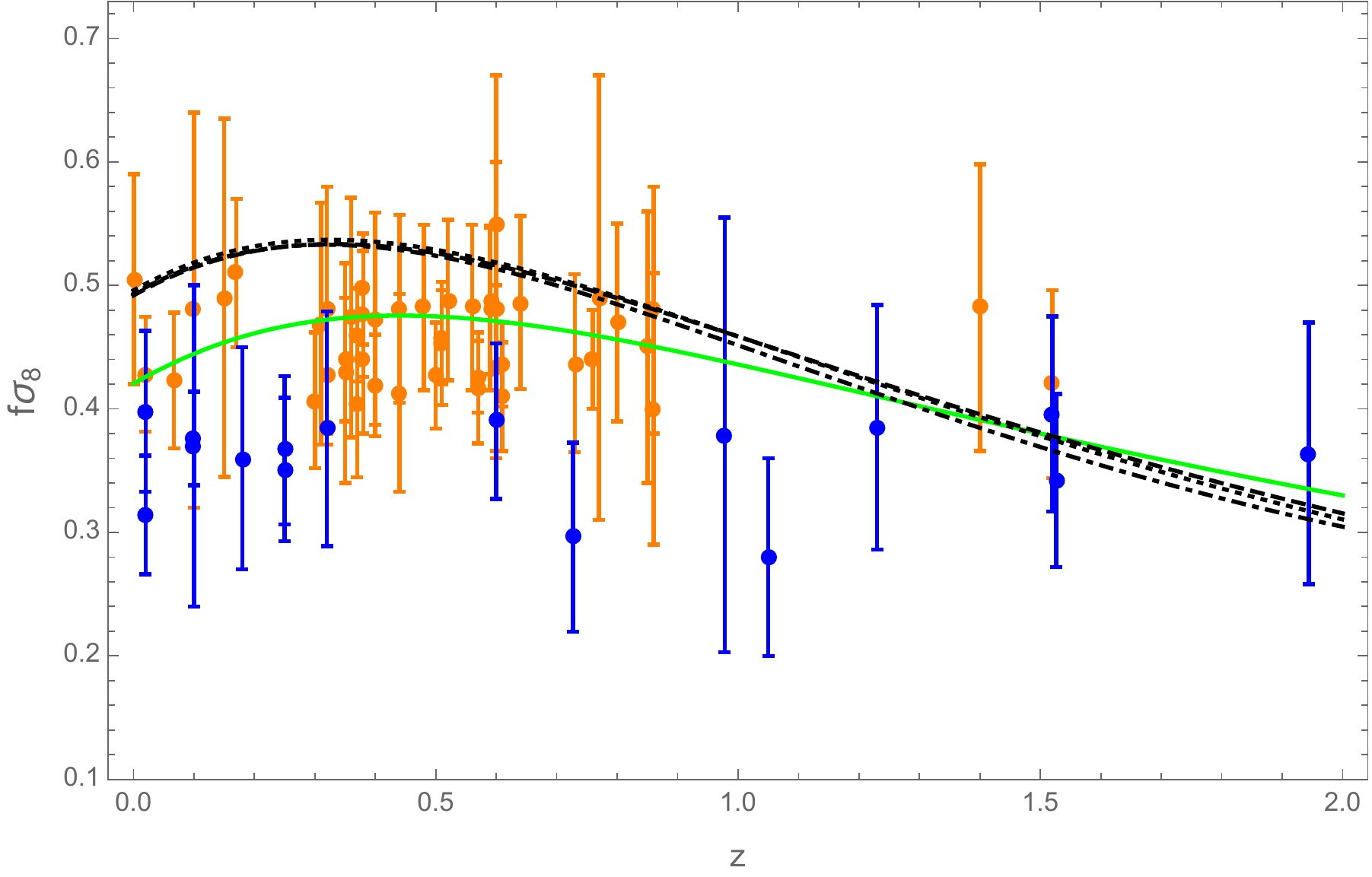}
\includegraphics[scale=0.6]{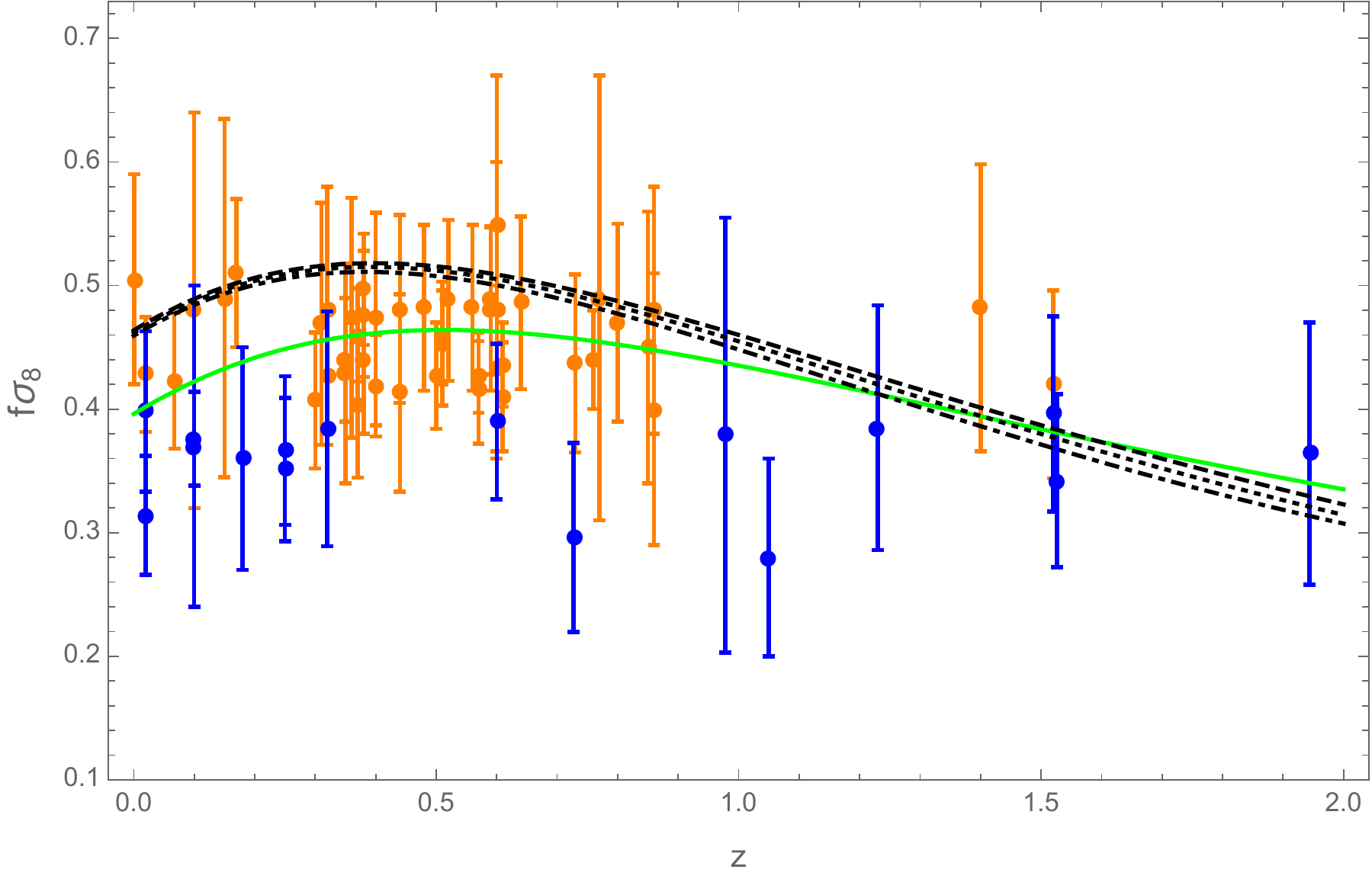}
\caption{The evolution of $f\sigma_8$ for the model (\ref{model2}) with $\lambda_1>1$. The different curves are calculated for the mode $k=300 a_0H_0$, assuming $\lambda_1=5\times 10^2, 5\times 10^3, 5\times 10^4 $ for $\eta=3,4,5$ respectively. The upper Fig. corresponds to $\Omega_{m0}=0.30$ and for the lower curves we used $\Omega_{m0}=0.27$. The curves move towards values greater than $\Lambda$CDM-curve for $z<1.3$, worsening the $\sigma_8$-tension.}
\label{fig10}
\end{center}
\end{figure}

\newpage
\section{Discussion}
The exponential correction to GR introduced in the model (\ref{model2}) gives a viable DE model in the context of modified gravity. 
The model (\ref{model2}) satisfies the conditions of stability during the whole cosmological evolution from matter era to future de Sitter attractor, and passes local and cosmological viability tests. The background evolution of the main cosmological parameters is consistent with current observations, as shown in the numerical analysis illustrated in Figs. 1 and 2.\\
The chameleon mechanism for this model was analyzed and it was found that the model can satisfy the thin shell condition, that in turn allows spherically symmetric solutions with post-Newton parameter $\gamma$ close to $1$ with very high precision (such as $|\gamma-1|\sim 10^{-22}$ for $\eta=5,\lambda_1=10^4$), and therefore,  deviations from GR become highly suppressed. 
From table I it is observed that the deviation parameter $m(r)$ for $\eta>1$ grows faster from the matter era to the current epoch than in models with $\eta<1$, which has important consequences in the evolution of matter perturbations.\\
Analyzing the effect of the model on matter perturbations, It was found that for the wave number interval relevant to the linear regime in galaxy power spectrum, if the transition to scalar-tensor regime occurs at a time close to the current, then the deviation parameter for the model (\ref{model2}) may satisfy the bound $m(z\approx 0)\ge 5\times 10^{-6}$. If the transition to scalar-tensor regime occurred during deep matter era, then it affects the matter power spectrum compared to the $\Lambda$CDM on scales relevant to galaxy power spectrum. Numerical results in table II show that the bound $\Delta n(t_{\Lambda})<0.05$ is satisfied for $\eta>3$. \\
The growth index of matter density perturbations was analyzed for the cases $\lambda_1=1$ and $\lambda_1>1$ fro different values of $\eta$. For $\lambda_1=1$ and $\eta\ge 3$ the models are indistinguishable from $\Lambda$CDM in the background evolution, but they don't leave a visible pattern in the evolution of matter perturbations compared to $\Lambda$CDM. The main difference is present in the growth index, presenting dispersion for $3\le\eta<6$, which disappears for $\eta\ge 7$ (see Fig. 8) where the model becomes scale invariant and degenerate with respect to $\eta$. In principle this degeneracy could be broken if an accurate value for $\Delta n(t_{\Lambda})$ were known.
 In the cases considered for $\lambda_1>1$ we used the quantity $q=\lambda_1/y_{ds}^{\eta}$ and giving values to it of the order of $10^{-2}-10^{-3}$. It was found that the maximum of $f$ decreases and the transition redshift moves to lower values as $\eta$ increases. In all cases the domain of values of $\gamma(z)$ are far lower than $\gamma_0$ and the dispersion of $f(z\approx 0)$ is very small, which is also true for $\gamma(z\approx 0)$ except for the larger mode. It was also found that $\gamma(z\approx 0)$ is independent of $\eta$. \\
The weighted growth rate $f\sigma_8(z)$ provided by different galaxy surveys has becoming an important test for DE models, even more when these observations show tension with Planck 2015/$\Lambda$CDM data. Numerical analysis performed with the model (\ref{model2}), illustrated in Figs. 8 and 9, shows that at most the model aligns with the $\Lambda$CDM model for the parameters $\lambda_1=1$ and $\eta>4$. So the model does not contribute to lowering the tension between CMB and LSS observations. \\
Future observations of CMB and LSS may narrow the parameter space for the observables from matter perturbations, including $f\sigma_8$, which will be crucial in 
deciding on a dynamical DE model instead of the rigid cosmological constant, while reducing the number of potential candidates.

\section*{Acknowledgments}
This work was supported by Universidad del Valle and MINCIENCIAS - COLOMBIA, Grant No. 110685269447 RC-80740-465-2020, projects 69723 and 69553.

\end{document}